\documentclass[11pt,reqno]{amsart}

\usepackage[colorlinks,citecolor=blue,urlcolor=blue,linkcolor=blue]{hyperref}
\usepackage{amsmath}
\usepackage{amssymb}
\usepackage{bm}
\usepackage{graphicx}
\usepackage[latin1]{inputenc}
\usepackage{amsfonts}
\usepackage{amsthm}
\usepackage{newlfont}
\usepackage{natbib}
\usepackage{latexsym}
\usepackage{float}
\usepackage{capt-of}\usepackage{stackrel}
\usepackage{mathrsfs}
\usepackage{multirow}
\usepackage{xr}
\usepackage{tikz}
\usetikzlibrary{shapes.geometric, intersections}

\newlength{\defbaselineskip}
\newcommand{\setlinespacing}[1]{\setlength{\baselineskip}{#1\defbaselineskip}}

\numberwithin{equation}{section}
\newtheorem{theorem}{Theorem}[section]
\newtheorem{corollary}[theorem]{Corollary}

\newtheorem{prop}[theorem]{Proposition}
\newtheorem{ass}[theorem]{Assumption}

\newtheorem{defn}[theorem]{Definition}

\theoremstyle{remark}
\newtheorem{remark}[theorem]{Remark}
\theoremstyle{remark}

\DeclareFontFamily{U}{mathx}{}
\DeclareFontShape{U}{mathx}{m}{n}{<-> mathx10}{}
\DeclareSymbolFont{mathx}{U}{mathx}{m}{n}
\DeclareMathAccent{\widecheck}{0}{mathx}{"71}

\newcommand{\ml}{\lambda}
\newcommand{\argmax}{\operatorname{argmax}}

\newcommand{\rank}{\operatorname{rank}}
\newcommand{\col}{\operatorname{col}}
\newcommand{\miff}{\Leftrightarrow}

\newcommand{\BC}{\mathbb{C}}
\newcommand{\BR}{\mathbb{R}}
\newcommand{\BN}{\mathbb{N}}
\newcommand{\BZ}{\mathbb{Z}}

\newcommand{\wh}[1]{\widehat{#1}}

\newcommand{\conv}{\rightarrow}
\newcommand{\toinf}{\conv\infty}
\newcommand{\tozero}{\conv 0}
\newcommand{\toone}{\conv 1}
\newcommand{\convp}{\stackrel{p}{\conv}}

\newcommand{\convw}{\stackrel{w}{\conv}}

\newcommand{\V}{\mathbb{V}}
\newcommand{\setB}{\mathcal{B}}
\newcommand{\setC}{\mathcal{C}}

\newcommand{\E}{\mathbb{E}}
\newcommand{\myH}{\mathscr{H}}
\newcommand{\mysingle}{single}
\newcommand\myPr{\mathbb{P}}
\newcommand{\SR}{\mathbb{R}}

\newcommand{\rd}{\ensuremath{\mathrm{d}}}

\newcommand{\tref}[1]{$\ref{#1}$}
\newcommand{\cca}{\texttt{cca}}

\def\cadlag{c\`adl\`ag}
\newcommand{\KL}{Karhunen-Lo\`eve}
\newcommand{\KLB}{$KL$ basis}
\newcommand{\myiid}{\text{i.i.d.}}
\newcommand{\myR}{\mathcal{R}}

\newcommand{\myA}{\mathcal{A}}
\newcommand{\cs}{cointegrating space}
\newcommand{\as}{attractor space}

\newcommand{\indB}{z}
\newcommand{\indC}{w}
\newcommand{\indD}{v}

\newcommand{\myHs}{\myH_0^*}
\newcommand{\myHss}{\myH_1^*}

\begin{document}

\setlinespacing{1.5}

\title[]{Inference on the attractor space via functional approximation}

\author[]{Massimo Franchi, Paolo Paruolo\\\\\today}
\thanks{\\\emph{Acknowledgments}: The views expressed in the paper are those of the authors and do not necessarily reflect the opinion of the institutions of affiliation. The authors thankfully acknowledge useful comments from
two anonymous referees,
Sophocles Mavroeidis,
Anindya Banerjee,
Peter Boswjik,
David Hendry,
S\o ren Johansen,
and participants to the
``Workshop to celebrate 40 years of Unit Roots and Cointegration'' 7-9 April 2025, Oxford,
and the ``Eleventh Italian Congress of Econometrics and Empirical Economics'',
 29-31 May 2025, Palermo. M. Franchi gratefully acknowledges partial financial support from PRIN 20223725WE ``Methodological and computational issues in large-scale time series models for economics and finance''.
\\
\indent
 M. Franchi, Sapienza University of Rome,
\href{https://orcid.org/0000-0002-3745-2233}{ORCID: 0000-0002-3745-2233}
 E-mail: massimo.franchi@uniroma1.it,
 Address: P.le A. Moro 5, 00185 Rome, Italy.
 \\\indent
 P. Paruolo.
 European Commission, Joint Research Centre (JRC)
\href{https://orcid.org/0000-0002-3982-4889}{ORCID: 0000-0002-3982-4889}, Email: paolo.paruolo@ec.europa.eu,
 Address: Via Enrico Fermi 2749, 21027 Ispra (VA), Italy.
 }

\begin{abstract}
This paper discusses semiparametric inference on hypotheses on the cointegration and the attractor spaces for $I(1)$ linear processes with moderately large cross-sectional dimension. The approach is based on
empirical canonical correlations and functional approximation of Brownian motions, and it can be applied both to the whole system and or to any set of linear combinations of it. The hypotheses of interest 
are cast in terms of the number of stochastic trends in specified subsystems, and inference 
is based either on selection criteria or on sequences of tests. This paper derives the limit distribution of these tests in the special one-dimensional case, and discusses asymptotic properties of the derived inference criteria
for hypotheses on the \as\
for sequentially diverging sample size and number of basis elements in the functional approximation.
Finite sample properties are analyzed via a Monte Carlo study and an empirical illustration on exchange rates is provided.
\end{abstract}

\keywords{Unit roots; cointegration; $I(1)$; \as; \cs; semiparametric inference; canonical correlation analysis\\ JEL Classification: C32, C33}

\begingroup
\def\uppercasenonmath#1{} 
\let\MakeUppercase\relax 
\maketitle
\endgroup

\section{Introduction}

\label{sec_intro}

The notion of cointegration introduced in \citet{EG:87} and its link to the existence of an Equilibrium Correction Mechanism (ECM), see \citet{DHSY:78}, have greatly influenced the econometric analysis of multiple time series. The introduction in \citet{Joh:88, Joh:91} of the Quasi Maximum Likelihood Estimator (QMLE) of the ECM within a Vector Autoregressive (VAR) model for variables integrated of order 1 (I(1)) provided a new set of  inferential tools that shaped the understanding of the vector-space nature of the \cs.

Hypotheses on the \cs\  were introduced together with the QMLE in the I(1) VAR case. Specifically, \citet[][eq. (3.1)]{Joh:91} considered the hypothesis that the \cs\ is contained in a pre-specified linear subspace, and also considered similar hypotheses on the matrix of adjustment coefficients. Soon after, \citet[][eq. (15)]{JJ:92} considered the converse hypothesis that the \cs\ contains a pre-specified linear subspace; see \citet[][Chapter 7.2]{Joh:96} for a summary.

These hypotheses on the \cs\ are often used to investigate economic-theory implications
after determination of the cointegration rank and prior to the identification of the cointegrating matrix, see \citet{Joh:96}, \citet{Jus:06} and Section \tref{sec_motivation} below for some examples. Inclusion restrictions of the same type for the set of linear combinations defining the innovations to the stochastic trends were also discussed in \citet[][]{GG:95} and subsequent literature.
The present paper focuses on this type of hypotheses on the \as, which is the complementary orthogonal space with respect to the \cs.

Recent years have seen the surge in availability of datasets with cross-sectional dimension $p$ in the double digits, coupled with a time series dimension $T$ in the double or triple digits; several contributions have recently addressed this case. One approach focuses on the VAR model with $p$ and $T$ diverging proportionally: \citet{OW:18,OW:19} characterize the distribution of the likelihood ratio test of \citet{Joh:88,Joh:91} for the null of no cointegration, \citet{BG:22,BG:24} propose a modified version of the likelihood ratio test statistics for the null of no cointegration, and \citet{LS:19} discuss Lasso-type estimators of the number of stochastic trends.

Other recent contributions have been cast in a semiparametric framework that encompasses but it is not restricted to VAR processes, see for instance \citet{Bie:97}, \citet{ChPh:09}, \citet{BT:22}, \citet{BCT:24}, \citet{FGP:23,FGP:24} and references therein. Related approaches based on a fixed number $K$ of cosine functions are presented in \citet{MW:08,MW:17,MW:18}, who respectively estimate the number of stochastic trends and the cointegrating relations using an IV approach, and analyze long-run variability and covariability of filtered time series.

The present paper considers inclusion restrictions between the \as\ and a given linear subspace, within the semiparametric context of \citet{FGP:24}, henceforth FGP, with $T$ diverging and $p$ fixed. The tools proposed in this paper cover the case of moderate cross section dimension $p$ in the sense of \citet{CS:24}, and can be used together with the estimators proposed in FGP for the estimation of the cointegration rank and of a basis of the \as.

The contributions of this paper include: (i) the definition of the hypotheses of interest as hypotheses on the number of stochastic trends on subsets of variables; (ii) the characterization of the restricted parameter space under the null and alternative hypotheses;
(iii)
the derivation of decision rules for these hypotheses both in terms of selection criteria for the number of stochastic trends and in terms of sequences of tests of hypotheses; (iv) the derivation of exact formulae for the probability density function (p.d.f.) and the cumulative distribution function (c.d.f.)
of the limit distribution for the tests of hypothesis in the univariate case; (v) a Monte Carlo study of the finite sample properties of the proposed procedures and an application to a system of $p=20$ exchange rates.

Following FGP, this paper considers a $p$-dimensional I(1) linear process $\Delta X_t=C(L)\varepsilon_t$ with Beveridge Nelson decomposition $C(z)=C+(1-z)C_1(z)$ with $s:=\rank C$, see e.g. \citet{PS:92}. The integer $s$ is the number of
Common Trends (CT) and if $s>0$, $C$ has representation $C=\psi \kappa'$ for $\psi, \kappa$ of full column rank $s$, and $X_t$ satisfies the CT representation
\begin{equation}\label{eq_Xfactor}
X_t=\gamma+\psi\kappa'\sum_{i=1}^t\varepsilon_i +C_1(L)\varepsilon_t,\qquad t=1,2,\dots,
\end{equation}
where $L$ (respectively $\Delta= 1-L$) is the lag (respectively difference) operator, $\gamma$ is a vector of initial values, $\varepsilon_t$ is a vector white noise, $\kappa'\sum_{i=1}^t\varepsilon_i$ are $0 < s\leq p$ linearly independent random walks (stochastic trends), $\psi$ is a $p\times s$ full column rank loading matrix spanning the \as\ $\col\psi$, and $C_1(L)\varepsilon_t$ is an $I(0)$ linear process; here $C(L)$ and  $C_1(L)$ are infinite matrix polynomials and $\beta$ is a basis of the \cs\ $(\col\psi)^\bot$, the orthogonal complement of $\col\psi$. When $s=0$, $X_t$ does not contain stochastic trends. 

This paper considers hypotheses $\col b \subseteq \col \beta \subseteq \col B $ that can be reformulated as
$\col a \subseteq \col \psi \subseteq \col A$ in terms of the \as,
with $a=B_\bot$, $\psi=\beta_\bot$, $A=b_\bot$, and it uses estimators of $s$ introduced in FGP.
The present study constructs inference criteria for the validity of restrictions of the type $\col a \subseteq \col \psi $ or $\col \psi \subseteq \col A $, as well as appropriate asymptotic tests of these hypotheses.

The approach is based on the empirical canonical correlations between the $p\times 1$ vector of observables $X_t$ and a $K\times 1$ vector of deterministic variables $d_t$, constructed as the first $K$ elements of an orthonormal $L^2[0,1]$ basis discretized over the equispaced grid $1/T,2/T,\dots,1$. In the asymptotic analysis, the cross-sectional dimension $p$ is fixed while $T$ and $K$ diverge sequentially, $K$ after $T$, denoted by $(T,K)_{seq}\toinf$.

The proposed inferential procedures are based on the \emph{dimensional coherence} of the semiparametric approach, which guarantees that assumptions that hold for the whole of $X_t$ also coherently apply to linear combinations of $X_t$, see \cite{JoJu:14} for a discussion of this concept in VARs. Dimensional coherence is used here to address inference on the cointegration and attractor subspaces.

The rest of the paper is organized as follows. The rest of this section introduces notation,
Section \ref{sec_motivation} contains motivating examples of hypotheses on the attractor space;
Section \ref{sec_setup} discusses the assumptions on the Data Generating Process (DGP);
Section \ref{sec_hyp} discusses alternative formulations of the hypotheses of interest;
Section \ref{sec_main} presents the proposed inference procedures;
Section \ref{sec_asy} discusses their asymptotic properties;
Section \ref{sec_M} contains a Monte Carlo experiment;
Section \ref{sec_app} reports an empirical application to exchange rates and Section \ref{sec_conc} concludes. Appendix \ref{sec_proofs} contains proofs.

\subsection*{Notation} The following notation is employed in the rest of the paper. For any matrix $a \in \BR^{p\times q}$, $\col a$ indicates the linear subspace of $\BR^{p}$ spanned by the columns of $a$; for any full column rank matrix $a$, the orthogonal projection matrix onto $\col a$ is $\bar a a'=a \bar a'$, where $\bar a:=a(a'a)^{-1}$, and $a_\bot$ of dimension $p\times (p-q)$ denotes a matrix whose columns span the orthogonal complement of $\col a$, i.e. $\col a_\bot = (\col a)^\bot$. 
The symbol $e_j$ indicates the $j$-th column of the identity matrix $I$ and $\iota$ indicates a column vector of ones.

For any $a \in \BR$, $\lfloor a \rfloor$ and $\lceil a \rceil$ denote the floor and ceiling functions; for any condition $c$, $1(c)$ is the indicator function of $c$, taking value 1 if $c$ is true and 0 otherwise. 
The space of right continuous functions $f:[0,1]\to\BR^n$ having finite left limits, endowed with the Skorokhod topology, is denoted by $D_n[0,1]$, see \citet[][Ch. VI]{JS:03} for details, also abbreviated to $D[0,1]$ if $n=1$.
A similar notation is employed for the space of square integrable functions $L_n^2[0,1]$, abbreviated to $L^2[0,1]$ if $n=1$.
Riemann integrals $\int_0^1 a(u)b(u)'\rd u$,
are abbreviated
as $\int ab'$.
Finally the complement of set $\mathcal{A}$ is indicated as $\mathcal{A}^\complement$.

\section{Motivation}\label{sec_motivation}
This section provides examples of hypotheses on the attractor space related to aggregation, balanced growth and the autonomy of trends for subsets of variables; these examples motivate interest in hypotheses on the \as. Here $X_t$ is assumed to satisfy \eqref{eq_Xfactor} with $s$ common stochastic trends.

\subsection{Aggregation}
Consider the following aggregation problem. Assume $X_{it}$ represents value-added in region $i$ in a given country, where $i\in \mathcal{I}:=\{1,\dots,p\}$ and $X_{t}=(X_{1t},\dots, X_{pt})'$.
Let also $X_{t}$ be $I(1)$; this means that value-added is stochastically trending, at least in some region; for simplicity, assume first that all $X_{it}$ are $I(1)$.

One is interested in an aggregate or average value-added in the country $w'X_{t}$, where $w=p^{-1}\iota$. 
Ideally one would like $w'X_{t}$ to share the same trend with all the regional components $X_{it}$; in this case, in fact, the aggregate $w'X_{t}$ would have the same information content -- in terms of trends -- as each $X_{it}$; call this property `aggregation invariance'.

This requirement imposes restrictions on the \as\ in DGP \eqref{eq_Xfactor}, and provides an example of hypothesis that can be tested with the inference tools developed in this paper. In fact, observe that $w'X_{t}$ and $X_{it}$ have the same trend whenever $(w-e_{i})'X_{t}$ is not trending, i.e. when $v_i:=w-e_{i}$ is a cointegrating vector.
Hence the hypothesis of aggregation invariance requires that the number of stochastic trends $s$ in \eqref{eq_Xfactor} is at least equal to 1 (in order for $X_{t}$ to be I(1)) and that $\mathcal{V}:=\col(v_{1},\dots, v_{p}) \subseteq \col \beta$ with $v_{i}:=w-e_{i}$.

Consider for instance $p=3$, and $\mathcal{V}=\col(v_{1},v_{2},v_{3})$ with $v_1=\frac13(-2,1,1)'$,
$v_2=\frac13(1,-2,1)'$, $v_3=\frac13(1,1,-2)'$. It is simple to note that $\rank(v_{1},v_{2},v_{3})=2$
because $v_1+v_2+v_3=0$, i.e. $\dim \mathcal{V}=p-1$,
and that
$\col(v_{1},v_{2},v_{3})=\col b$, where $b=(b_1,b_2)$ can be chosen as $b_1:=(v_2-v_1)=(1,-1,0)'$, and $b_2:=(v_3-v_1)=(1,0,-1)'$.\footnote{Of course, other choices of $b$ are also possible, such as for instance $
b=(v_2-v_1, v_3-v_2)$.}
Note that $\col b^{\bot} = \col \iota$, and this implies that $\col(b) \subseteq \col \beta$ corresponds to $\col \psi \subseteq \col \iota$. This derivation can be easily written for generic $p$; aggregation invariance hence implies $\col \psi \subseteq \col \iota$.

Because $s:=\dim\col\psi\geq 1$ and $\dim \col \iota= 1$, the inclusion $\col \psi \subseteq \col \iota$ needs to hold as an equality, i.e. $\col \psi = \col \iota$ and $s=1$, and hence one can choose $\psi = \iota$. This shows that aggregation invariance corresponds to a special type of inclusion restriction of the \as, of the form $\psi =\iota$.

Proposition \ref{prop_lin_aggr}.(a) below shows that this result extends to a more general situation where one considers any weighted sum $w'X_t$ of elements in $X_t$, where $w$ is a nonzero vector with $w'\iota$ normalized to 1, $w'\iota=1$. Part (b) discusses the case where the regional component $X_{it}$ is non-trending for $i$ in $\mathcal{I} \setminus \mathcal{I} _1$, and shows that the same result holds restricting the elements of $\psi$ to be 0 for the non-trending regions.

\renewcommand{\theenumi}{(\roman{enumi})}
\renewcommand{\labelenumi}{\theenumi}

\begin{prop}[Aggregation invariance]\label{prop_lin_aggr}
Let $w$ be a given nonzero vector with $w'\iota=1$. \\ 
\textnormal{(a)} The following statements are equivalent:
\begin{enumerate}
\item 
$ w'X_{t}=I(1)$ and $X_{it}=I(1)$ share the same common trend for any $i\in \mathcal{I}$;
\item $\psi =\iota$ with $s=1$ in \eqref{eq_Xfactor}.
\end{enumerate}
\textnormal{(b)} The same equivalence between \textnormal{(i)} and \textnormal{(ii)} holds when substituting $\mathcal{I}$ with any of its nonempty subsets $\mathcal{I}_1$ replacing $\psi=\iota$ with
$\psi=\sum_{i \in \mathcal{I}_1}e_i$.
\end{prop}

This example shows that some hypotheses regarding aggregation of trending series translate into hypotheses on $\col \psi$ of the type studied in this paper.

\subsection{Balanced growth}
Another hypothesis of interest concerning the \as\ arises when analyzing the balanced growth of different countries, see \citet{KPSW:91}. For simplicity consider the case of 2 countries, with
$(X_{1t},X_{2t},X_{3t})'$ (respectively $(X_{4t},X_{5t},X_{6t})'$) representing log income, log consumption and log investment in county 1 (respectively country 2).
Balanced growth in a given country means that log income, log consumption and log investment share the same common trend. This implies that the number of trends $s$ must
be either 1 or 2, depending on whether the 2 countries are on the same trend path or on different ones.

Assume first that $s=2$.
One may wish to test that each of the two countries is on its own balanced growth path.
Balanced growth in the first country corresponds to the hypothesis $\col a_1 \subseteq \col \psi$ with
$a_1 :=\sum_{i=1}^{3}e_i = (1,1,1,0,0,0)'$, while balanced growth in the second country is given by $\col a_2 \subseteq \col \psi$ with
$a_2 :=\sum_{i=4}^{6}e_i=(0,0,0,1,1,1)'$. These two hypotheses could be tested separately as $\col a_i \subseteq \col \psi$, $i=1,2$, or also jointly by considering $\col(a_1,a_2) = \col \psi$. These are three examples of hypotheses of the type $\col a \subseteq \col \psi$ considered in this paper.

Assume now that $s=1$. Perfectly balanced growth between the two countries corresponds to the hypothesis $\psi=\iota$; under this assumption, country 1 and 2 are on the same growth path with equal loadings (same scale of the trend). Alternatively, proportional balanced growth can be represented as $\psi = (1,1,1,c,c,c)'$ with 
$c\neq 0,1$
representing the ratio between the loadings of country 2 and 1.
Under this hypothesis, country 1 and 2 are on the same growth path but with different loadings (and hence different scale of the trend). This is an example of an hypothesis of the type $\col \psi \subseteq \col A$, where $A=(a_1,a_2)$.

\subsection{Autonomous trends for subsets}\label{sec_sub_autonomy}
In various applications, including the one in Section \ref{sec_app}, one is interested in testing whether a subset of variables $X_{it}$ does not cointegrate, with $i \in \mathcal{J} \subset \mathcal{I}:=\{1,2,\dots,p\}$. This hypothesis can be expressed as $\col \{e_i, i\in \mathcal{J}\} \subseteq \col \psi$, an hypothesis of the type $\col a \subseteq \col \psi$.
All of the above are examples of hypotheses discussed in this paper.

\section{Data Generating Process}\label{sec_setup}
This section introduces the assumptions on the DGP, the property of dimensional coherence and the hypotheses of interest.
\subsection{Assumptions}
Let $\{X_t\}_{t\in \BZ}$ be a $p\times 1$ linear process generated by
\begin{equation}\label{eq_DX}
\Delta X_t=C(L)\varepsilon_t,
\end{equation}
where $L$ is the lag operator, $\Delta:=1-L$, $C(z)=\sum_{n=0}^\infty C_n z^n$, $z\in \BC$, satisfies Assumption \tref{ass_DGP} below, and the innovations $\{\varepsilon_t\}_{t\in \BZ}$ are independently and identically distributed (i.i.d.) with expectation $\E(\varepsilon_t)=0$, finite moments of order $2+\epsilon$, $\epsilon>0$, and positive definite variance-covariance matrix $\V(\varepsilon_t)=\Omega_\varepsilon$, indicated as $\varepsilon_t\:\myiid\: (0,\Omega_\varepsilon)$, $\Omega_\varepsilon>0$.\footnote{The i.i.d. assumption can be relaxed to a martingale difference sequence; this is not done here for simplicity.}

\begin{ass}[Assumptions on $C(z)$]\label{ass_DGP}
Let $C(z)=C+C_1(z)(1-z)$ in \eqref{eq_DX} be of dimension $p\times n_\varepsilon$, $n_\varepsilon \geq p$, and satisfy the following conditions:
\begin{enumerate}
\item[(i)] $C(z)=\sum_{n=0}^\infty C_n z^n$ converges for all $|z|<1+\delta$, $\delta >0$;
\item[(ii)] $\rank C(z)<p$ only at isolated points $z=1$ or $|z|>1$;
\item[(iii)] $\psi_\bot'C_1\kappa_\bot$ has full row rank $r:=p-s$,
with $C_1:=C_1(1)$, $\psi_\bot$ $($respectively $\kappa_\bot)$ is a basis of $(\col C)^\bot$ $($respectively $(\col C')^\bot)$ and $s:=\rank C$.
\end{enumerate}
\end{ass}

Condition (i) is a standard one for integrated linear processes of integer order, see \citet{Joh:96}. In the case of $p=n_\varepsilon$ conditions (ii) and (iii) are necessary and sufficient conditions for the process to have a (infinite order) VAR representation with error correction, see \citet{Joh:96}, Chapter 4. In the present context, condition (iii) guarantees that the relevant long run variance is nonsingular, see \eqref{eq_x} and \eqref{eq_D_fullrank} in the Appendix.
Cumulating \eqref{eq_DX}, one finds the Common Trends (CT) representation
\begin{equation}\label{eq_X}
X_t=\gamma+ \psi \kappa'\sum_{i=1}^t\varepsilon_i +C_1(L)\varepsilon_t,\qquad t=1,2,\dots,
\end{equation}
which gives a decomposition of $X_t$ into initial conditions $\gamma:=X_0-C_1(L)\varepsilon_0$, the $s$-dimensional random walk component $ \kappa'\sum_{i=1}^t\varepsilon_i$ with loading matrix $\psi$ and the I(0) linear process $C_1(L)\varepsilon_t$. Initial conditions can be treated as in \citet{Joh:96} or as in \citet{Ell:99}, see also \citet{MW:08}.

Observe that $X_t$ is $I(0)$ if $s=0$, i.e. $r=p$; it is $I(1)$ and cointegrated if $0<r,s<p$ and it is $I(1)$ and non-cointegrated if $s=p$, i.e. $r=0$. Assumption \tref{ass_DGP} includes possibly nonstationary VARMA processes (for $n_\varepsilon = p$) in line with e.g. \citet{SW:88} and Dynamic Factor Models (for $n_\varepsilon > p$) as in e.g. \citet{Bai:04}.

If $s>0$, the matrix $\psi$ in \eqref{eq_X} is $p \times s$ with columns that form a basis of $\col C$ called the {\as} of $X_t$. Its orthogonal complement $(\col C)^\bot$ is the {\cs} of $X_t$, and the columns of $\beta:=\psi_\bot$ (of dimension $p \times r$) form a basis of the \cs.

\subsection{Dimensional coherence}

The assumptions on the DGP of $X_t$ imply that they also apply to linear combinations of $X_t$. In fact
whenever $X_t$ has a CT representation \eqref{eq_X}, linear combinations of $X_t$ also admit a CT representation, with a number of stochastic trends that is at most equal to that in $X_t$. 

FGP show that when
$H$ is a $p\times m$ full column rank matrix and $X_t$ satisfies \eqref{eq_DX} with $C(z)$ fulfilling Assumption \tref{ass_DGP}, then the same holds for $H'X_t$, i.e. $\Delta H'X_t=G(L)\varepsilon_t$, where $G(z):=H'C(z)$ satisfies Assumption \tref{ass_DGP} with $G(1)=H'C$ of rank $j\leq s$.
This property is called `dimensional coherence' of Assumption \tref{ass_DGP}.

\subsection{Hypotheses of interest}
The hypotheses of interest are formulated in terms of $\beta$ as $\col b \subseteq \col \beta \subseteq \col B$, or equivalently in terms of $\psi$ as
\begin{equation}\label{eq_hyp}
\col a \subseteq \col \psi \subseteq \col A,
\end{equation}
where $a=B_\bot$, $\psi=\beta_\bot$, $A=b_\bot$, and $b$ and $B$ are full column rank matrices. The aim of the paper is to make inferences about
\eqref{eq_hyp}
from a sample $\{X_t\}_{t = 0,1,\dots,T}$ observed from \eqref{eq_X}.

\section{Formulation of the hypotheses}\label{sec_hyp}

This section discusses the hypotheses of interest and their implications on the DGP. It also relates them to the dimensional coherence property in Section \tref{sec_setup}.

Let $X_t$ be generated as in \eqref{eq_X} with $s$ stochastic trends, $s>0$ and let $\Psi:=\{\psi \in \SR^{p \times s},\rank \psi = s\}$ indicate the unrestricted parameter space for $\psi$, i.e. the set of all DGPs that satisfy \eqref{eq_X} with stochastic trends loading matrix of rank $s$.
Consider hypotheses $\myH_0$ on the \cs\ $\col\beta $ or the \as\ $\col\psi $ of type \eqref{eq_hyp}, formulated as follows
\begin{align}\label{eq_H01a}
\myH_0
&: \:\:
\col b\subseteq \col \beta
\:\:
&\miff \:\:
\col\psi \subseteq \col A
\:\:
&\miff \:\:
\psi\in \Psi_0:=\{\psi = A \theta, \,\,\,\,\rank \theta = s\},
\\
\label{eq_H02a}
\myH_0
& : \:\:
\col \beta \subseteq \col B
\quad
&\miff \:\:
\col a\subseteq \col\psi
\:\:
&\miff \:\:
\psi\in\Psi_0:=\{\psi = (a, a_\bot \varphi), \,\,\,\, \rank \varphi = s-q\},
\end{align}%
where $A=b_\bot$ is $p\times m$, $\theta$ is $m \times s$, $a=B_\bot$ is $p\times q$, $a_\bot$ is $p\times (p-q)$, $\varphi$ is $(p-q) \times (s-q)$, all of full column rank, and $q\leq s\leq m$.\footnote{In the following, the convention is that empty matrices are absent, e.g. $\psi =a$ when $q=s$ in \eqref{eq_H02a}.}

The restricted parameter spaces $\Psi_0 \subset \Psi$ in \eqref{eq_H01a} and \eqref{eq_H02a} are next characterized in terms of
specific subsystems of variables, on which the dimensional coherence property applies.
Under hypothesis $\myH_0$ in \eqref{eq_H01a}, pre-multiplying \eqref{eq_X} by $(\bar{A},\bar{A}_\bot)'$
one finds%
\begin{align}\label{eq_H01A}
\bar A'X_{t} &=\theta \kappa '\sum_{i=1}^{t}\varepsilon
_{i}+\bar A'C_{1}(L)\varepsilon _{t}+\bar A'\gamma  \\ \label{eq_H01Aort}
\bar{A}_\bot'X_{t} &=\qquad \qquad \quad \bar{A}_\bot'C_{1}(L)\varepsilon _{t}+\bar{A}_\bot'\gamma
\end{align}%
where $\rank(\theta \kappa ')=s$. Because $X_t=(A,A_\bot)(\bar{A},\bar{A}_\bot)'X_{t}$, \eqref{eq_H01a} is equivalent to the fact that
(i) $A'X_t$ contains $s$ stochastic trends 
and (ii) $A_\bot'X_t$ contains $0$ stochastic trends.
Similarly, consider  hypothesis $\myH_0$ in \eqref{eq_H02a} and
partition $\kappa $ conformably with $\psi =(a,a_\bot \varphi)$ as $\kappa =:(\kappa_{1},\kappa _{2})$ with $q$ and $
s-q$  columns respectively, so that $\psi \kappa' =a \kappa_1' + a_\bot\varphi \kappa_2'$.
Pre-multiplying \eqref{eq_X} by $(\bar{a},\bar{a}_\bot)'$
one finds
\begin{align}\label{eq_H02_a}
\bar a ' X_{t} &=\kappa _{1} ' \sum_{i=1}^{t}\varepsilon
_{i}+\bar a ' C_{1}(L)\varepsilon _{t}+\bar a ' \gamma
\\ \label{eq_H02_aort}
 \bar a_\bot' X_{t} &=\varphi\kappa _{2}' \sum_{i=1}^{t}\varepsilon
_{i}+ \bar a_\bot' C_{1}(L)\varepsilon _{t}+ \bar a_\bot'\gamma
\end{align}%
where $\rank(\varphi\kappa _{2}')=s-q$. Hence \eqref{eq_H02a} is equivalent to the fact that
(i) $a'X_t$ contains $q$ stochastic trends
and (ii) $a_\bot'X_t$ contains $s-q$ stochastic trends.

Note that both hypotheses \eqref{eq_H01a} and \eqref{eq_H02a} have leading known subspace spanned by a $p \times l $ matrix $H$, where $(H,l)=(A,m)$ in \eqref{eq_H01a} and $(H,l)=(a,q)$ in \eqref{eq_H02a}. Let $n:=\min(l, s)$;
the next theorem shows the
formal equivalence between \eqref{eq_H01a} and \eqref{eq_H01A}, \eqref{eq_H01Aort} and between \eqref{eq_H02a} and \eqref{eq_H02_a}, \eqref{eq_H02_aort}.

\begin{theorem}[Parameter space decomposition under the null]\label{thm_equiv}
The restricted parameter space $\Psi_0$ under the null hypotheses \eqref{eq_H01a} or \eqref{eq_H02a} has the following representation
\begin{align}\label{eq_rest_par_space}
\myH_{0}: \psi \in \Psi_{0} = \Psi_{0 1}\, \cap \, & \Psi_{0 2}\\
\Psi_{0 1}:=\{\psi \in \Psi: \rank(H'\psi )=n\},
\quad & \quad \Psi_{0 2}:=\{\psi \in \Psi: \rank(H_{\bot}'\psi )=s-n\}, \nonumber
\end{align}%
where $(H,n)=(A,s)$ in \eqref{eq_H01a} and $(H,n)=(a,q)$ in \eqref{eq_H02a}.
\end{theorem}

In order to characterize the parameter space under the alternative $\myH_{1}: \psi \in \Psi_{0}^{\complement}:=\Psi \setminus\Psi_{0}$, where $\complement$ indicates the complement of a set,
it is useful to observe that $\rank\left( H'\psi \right) \leq n$; this means that
$\Psi_{01}$ in \eqref{eq_rest_par_space} requires $\rank\left( H'\psi \right)$ to be maximal, i.e. equal to $n$. Moreover,
consider the following inequality
\begin{equation}\label{eq_rank_ineq}
s:=\rank \psi=\rank((\bar{H},\bar{H}_\bot)(H,H_\bot)'\psi
)=\rank\left(
\begin{array}{c}
H'\psi  \\
H_\bot'\psi
\end{array}%
\right) \leq \rank\left( H'\psi \right) +\rank\left( H_{\bot
}'\psi \right)
\end{equation}%
which implies $\rank\left( H_{\bot
}'\psi \right) \geq s-\rank\left( H'\psi \right)$, and hence,
substituting
$\rank\left( H'\psi \right) \leq n $, one has
$\rank\left( H_{\bot
}'\psi \right) \geq s-n$. This means that
$\Psi_{02}$ in \eqref{eq_rest_par_space}
requires $\rank\left( H_\bot ' \psi \right)$ to be minimal, i.e. equal to $s-n.$
Summarizing, for any $\psi\in\Psi$ one has
$$
\rank\left( H'\psi \right) \leq n \qquad \rank\left( H_{\bot
}'\psi \right) \geq s-n,
$$
with equalities under the null hypothesis $\psi\in\Psi_0$, see \eqref{eq_rest_par_space}.
Under the alternative at least one of these inequalities is strict, i.e.
\begin{align}\label{eq_H1}
 \myH_{1}: \psi \in \Psi_{0}^{\complement} = \Psi_{01}^{\complement} & \, \cup \, \Psi_{02}^{\complement} \\
 \Psi_{01}^{\complement} = \{\psi \in \Psi: \rank(H'\psi )<n \}, \quad & \quad  \Psi_{02}^{\complement} = \{\psi \in \Psi: \rank(H_{\bot}'\psi )>s-n\}.\nonumber
\end{align}

Inequality \eqref{eq_rank_ineq} has also the following implications about the nesting between parameter spaces under the null $\myH_0$ and
under the alternative $\myH_1$.
\begin{theorem}[Nesting of subspaces and decomposition under the alternative]\label{thm_possible}
One has
\begin{enumerate}
\item $\Psi_{02}\subset \Psi_{01}$ $($i.e. $\Psi_{02}^{\complement}\supset \Psi_{01}^{\complement})$, 
\item $\Psi_{0}^{\complement} = \Psi_{02}^{\complement}$,
\item $\Psi_{0}^{\complement}=\Psi_{01}^{\complement} \cup (\Psi_{01}\cap \Psi_{02}^{\complement})$.
\end{enumerate}
\end{theorem}

\begin{figure}[tb]
\begin{tikzpicture}
\draw[thick] (-2,-1.5) rectangle (2.5,1.5)
(2.5,0) node[right]{$\Psi$};

\draw[thick,blue] (0,0) ellipse (1.5 and 1)
(1.5, 0) node[right]{$\Psi_{01}$};

\draw[thick,gray] (-.5,0) ellipse (.8 and 0.5)
(0.3,0) node[right]{$\Psi_{02}$};
\end{tikzpicture}
\caption{Representation the unrestricted ($\Psi$) and restricted ($\Psi_{0}=\Psi_{01} \cap \Psi_{02}$) parameter space  in terms of subsets.}\label{fig_balls}
\end{figure}

Part (i) of the theorem states that there are no parameter values in $\Psi$ for which $\Psi_{02}$ holds and $\Psi_{01}$ fails so that there is no intersection between $\Psi_{01}^{\complement}$ and $\Psi_{02}$; this is represented in Figure \ref{fig_balls}.

Part (ii) of the theorem states that the parameter space under the alternative $\Psi_{0}^{\complement}$ is equal to the one of the alternative to the second subhypothesis ($\Psi_{02}^{\complement}$) which in term of Figure \ref{fig_balls} is everything outside the grey ellipse. Part (iii) states that this region can be partitioned in 2 subregions: the one between the grey and blue ellipse ($\Psi_{01}\cap\Psi_{02}^{\complement}$) and the one outside the blue ellipse ($\Psi_{01}^{\complement}$).

As an illustration let
\begin{equation}\label{eq_psi_exmpl}
\psi =\left(
\begin{array}{cc}
1 & 0 \\
0 & 1 \\
0 & 1%
\end{array}%
\right)=(e_1,e_2+e_3)
\end{equation}
and consider the hypothesis $\col a \subseteq \col \psi$ with $a=e_{1}$, $s=2$ and $q=1$, which is satisfied in \eqref{eq_psi_exmpl}. This hypothesis is of type \eqref{eq_H02a} and states that the first variable does not cointegrate, i.e. that it is autonomous, see Section \tref{sec_sub_autonomy}.
One can illustrate decomposition \eqref{eq_rest_par_space}
calculating $\rank(a'\psi
)=\rank(1,0)=1=q$  and  $\rank(a_\bot'\psi
)=\rank\left(
\begin{array}{cc}
0 & 1 \\
0 & 1%
\end{array}%
\right) =1=s-q$; one can see that $\psi \in \Psi_{01}$ and $\psi \in \Psi_{02}$, as needs to be the case, see \eqref{eq_rest_par_space}.

Consider next the hypothesis $\col a \subseteq \col \psi$ with $a=e_{2}$, which is not satisfied in \eqref{eq_psi_exmpl}, so that $\psi \in \Psi_{0}^{\complement}$. One can investigate if $\psi \in \Psi_{01}^{\complement}$ or $\psi \in \Psi_{02}^{\complement}$ (or both); one has $\rank(a'\psi
)=\rank(0,1)=1=q$, which means $\psi \in \Psi_{0 1}$, and $\rank(a_{\bot
}'\psi )=\rank(I_{2})=2>s-q$, which means $\psi \in \Psi_{02}^{\complement}$. This illustrates that $\myH_{1}$
holds, with only one component of the alternative being true ($\psi \in \Psi_{01}$ and $\psi \in \Psi_{02}^{\complement}$).

Finally consider the hypothesis $\col a \subseteq \col \psi$ with  $a=e_{2}-e_{3}$
which is also not satisfied in \eqref{eq_psi_exmpl}, i.e. $\psi \in \Psi_{0}^{\complement}$. Again one can verify
if $\psi \in \Psi_{01}^{\complement}$ or $\psi \in \Psi_{02}^{\complement}$; one finds $%
\rank(a'\psi )=\rank(0,0)=0<q=1$ which means $\psi \in \Psi_{01}^{\complement}$
and $\rank(a_\bot'\psi
) =\rank\left(
\begin{array}{cc}
1 & 0 \\
0 & 2%
\end{array}%
\right) =2>s-q=1$, which
means $\psi \in \Psi_{02}^{\complement}$. In this case $\myH_{1}$ holds, with both
components of the alternative
being verified ($\psi \in \Psi_{01}^{\complement}$ and $\psi \in \Psi_{02}^{\complement}$).

This illustrates all possibilities, see Theorem \tref{thm_possible}. 



\section{Statistical analysis}\label{sec_main}

This section defines the selection criteria for hypotheses $\myH_0$ based on  the estimators of the number $s$ of stochastic trends (and the cointegrating rank $r$) proposed in FGP. The estimators are based on canonical correlation analysis, which is introduced first.

\subsection{Canonical correlations}\label{sec_cca_KL}

Let $\varphi_K(u):=(\phi_1(u),\dots,\phi_K(u))'$ be a $K \times 1$ vector function of $u \in[0,1]$, where $\phi_1(u),\dots,\phi_K(u)$ are the first $K$ elements of some fixed orthonormal \cadlag\ basis of $L^2[0,1]$. The leading example is the \KL\ basis for Brownian motion which corresponds to
$\phi_k(u)=\sqrt 2\sin \left((k-\frac12)\pi u \right)$,
see \eqref{eq_B_L2} and \eqref{eq_KLB} in the Appendix. Let $d_t$ be the $K\times 1$ vector constructed evaluating $\varphi_K(\cdot)$ at the discrete sample points $1/T,\dots, (T-1)/T, 1$, i.e.
\begin{equation}\label{eq_d}
d_t:=\varphi_K(t/T):=(\phi_1(t/T),\dots,\phi_K(t/T))',\qquad K\geq p,\qquad t=1,\dots,T.
\end{equation}

For a generic $p$-dimensional variable $y_t$ observed for $t=1,\dots,T$, the sample canonical correlation analysis of $y_t$ and $d_t$ in \eqref{eq_d}, denoted as $\cca(y_t,d_t)$,\footnote{The notation $\cca(y_t,d_t)$ is a shorthand for $\cca(\{y_t\}_{t=1}^T,\{d_t\}_{t=1}^T)$; similar abbreviations are used in the following.} consists in solving the following generalized eigenvalue problem, see e.g. \citet{Joh:96} and references therein,
\begin{equation}\label{eq_BEP}
\cca(y_t,d_t):\qquad |\ml M_{yy}-M_{yd}M_{dd}^{-1}M_{dy}|=0,\qquad M_{ij}:=T^{-1}\sum_{t=1}^Ti_tj_t'.
\end{equation}
This delivers eigenvalues $1\geq \ml_1\geq\ml_2\geq\dots\geq\ml_p \geq 0$ (the squared canonical correlations) and corresponding eigenvectors $v_1,v_2,\dots,v_p$ which are used to define canonical variates.

The next two sections present estimators of
the number of stochastic trends $s$, generically indicated by $\wh s$. When needed, the notation
$\wh s (y_t)$ is employed to indicate the dependence on the system of variables
$y_{t}$ used in the $\cca(y_t,d_t)$; $y_{t}$ equals $X_t$ when performing the estimation on the full system, while
$y_{t}$ equals
$H'X_t$ and $H_\bot'X_t$
corresponding to subsystems
\eqref{eq_H01A}, \eqref{eq_H01Aort} for $H=A$, and \eqref{eq_H02_a}, \eqref{eq_H02_aort} for $H=a$.

\subsection{Argmax estimators}\label{sec_criteria} The following definition reports estimators based on the maximization of functions of the squared canonical correlations. 

\begin{defn}[Argmax estimators $\wh s$ of $s$]\label{def_s_hat}
Define the maximal gap $($max-gap$)$ estimator
of $s$ as
\begin{equation}\label{eq_def_shat}
\underset{i\in \{0,\dots, p\}}{\argmax}(\ml_i-\ml_{i+1}),\qquad\ml_0:=1,\qquad\ml_{p+1}:=0,
\end{equation}
where $1\geq \ml_1\geq\ml_2\geq\dots\geq\ml_p \geq 0$ are the eigenvalues of $\textnormal{\cca}(X_t,d_t)$ and $X_t$.
Define also the following \emph{alternative argmax estimator}:
\begin{align}\label{eq_criteria}
&
\underset{i\in \{0,1,\dots, p\}}{\argmax}\:\frac{\prod_{h=1}^i\ml_h}{\prod_{h=i+1}^p\left(\frac{T}{K}\ml_h\right)}, 
\end{align}
where 
empty products are equal to  $1$, and 
uses is made of
the rates in eq. \textnormal{(4.3)} below Theorem \textnormal{4.1} in FGP for $(T,K)_{seq}\toinf$.
\end{defn}


\begin{remark}[Alternative criteria]
The criterion \eqref{eq_criteria} applies ideas in \cite{Bie:97} to the set of eigenvalues in $\cca(X_t,d_t)$. Alternative criteria inspired by \citet{AH:13} when applied to eigenvalues in $\cca(X_t,d_t)$ are discussed in FGP; they replace the function in the argmax in \eqref{eq_criteria} with
\begin{align}\label{eq_criteria_AH}
 \frac{\ml_i}{\ml_{i+1}} \qquad \text{or} \qquad 
 \frac{\log\left(1+\ml_i/\sum_{h=i+1}^p\ml_h\right)}{\log\left(1+\ml_{i+1}/\sum_{h=i+2}^p\ml_h\right)},
\end{align}
where the optimization is over the set of integers $0\leq i \leq p-1$ for the first function and $0\leq i \leq p-2$ for the second function. Because these criteria do not cover the values $p$ (and also  $p-1$ for the last function), they are not considered in the rest of the paper.

\end{remark}

\subsection{Test sequences}\label{sec_est_test}
This section
defines estimators $\wh s$ of the number of stochastic trends based on sequences of test similar to the one in \citet{Joh:96} for cointegration rank determination.
The procedure is described here for the full system $X_t$ in \eqref{eq_X} with $s$ stochastic trends, but can be similarly be applied to  subsystems $H'X_t$ and $H_\bot'X_t$ as in \eqref{eq_H01A}, \eqref{eq_H01Aort} for $H=A$, and \eqref{eq_H02_a}, \eqref{eq_H02_aort} for $H=a$.

Consider hypothesis $\myHs:s=i$ versus $\myHss:s<i$
at significance level $\eta$
with test statistics
$K \pi^2 \|\tau^{(i)}\|_h$ where $ \tau^{(i)}:=(1-\ml_i,1-\ml_{i-1},\dots,1-\ml_1)'$, $\|\cdot\|_h$ is the $h$-norm for vector and  $\ml_1\geq\ml_2\geq\dots\geq\ml_p$ are the eigenvalues of $\cca(X_t,d_t)$.
The two special cases of interest are $h=1,\infty$; the choice $h=1$ gives $\|\tau^{(i)}\|_1=\sum_{k=1}^i(1-\ml_k)$, while $h=\infty$ yields $\|\tau^{(i)}\|_{\infty}=1-\ml_i$; these expressions are similar to the trace and $\ml_{\max}$ statistics in \citet{Joh:96}. The critical values $c_{i,h,\eta}$ for these tests are discussed in Section \tref{sec_asympt} below.

The test sequence is defined as follows.
Start with the test of $\myHs:s=i$ for $i=p$; if the test does not reject, then $\wh s = p$, otherwise progress further. Test $\myHs:s=i$ for $i=p-1,p-2,\dots$ and proceed as before until $\myHs:s=k$, say, is not rejected; in this case
$\wh s = k$. Otherwise, all tests of $\myHs:s=i$, $i=p,\dots,1$ reject, and $\wh s = 0$.

\subsection{Decision rules on $\myH_0$}

This section describes decision rules for rejecting $\myH_0:\psi\in\Psi_0$, see \eqref{eq_rest_par_space}. The rules are expressed as functions of a generic estimator $\wh s$  of the number of stochastic trends defined in Sections \tref{sec_criteria} and \tref{sec_est_test}; remark that when $\wh s$ is based on test sequences as in Section \tref{sec_est_test}, one needs to specify the significance levels used in each test in the sequence.

The `joint decision rule' \emph{does not  reject}
$\psi \in \Psi_0$
 whenever $\indB$ equals 1, where
\begin{align}\label{eq_accept_H01}
& \indB := \indC\cdot \indD,
\qquad
\indC := 1(\wh s(H' X_{t})=n),
\qquad
\indD := 1(\wh s (H_\bot'X_{t})=n-s),
\end{align}
and $(H,n)=(A,s)$ for \eqref{eq_H01a} and $(H,n)=(a,q)$ for \eqref{eq_H02a}.

Note that $\indB$ checks both implications $ \psi \in \Psi_{01}$ and $ \psi \in \Psi_{02}$, via the indicators $\indC$ and $\indD$,
and rejects
$\psi \in \Psi_0 $ if \emph{either} $ \psi \in \Psi_{01}$ \emph{or} $ \psi \in \Psi_{02}$
is rejected, or both. An alternative procedure, called the `\mysingle\ decision rule', \emph{does not reject} $\psi \in \Psi_0$ if $\indD$ in \eqref{eq_accept_H01} equals 1, and rejects $\psi \in \Psi_0$ if $\indD$ equals 0. The rationale for this single decision rule is that if $\psi \in \Psi_{02}$, then also $\psi \in \Psi_{01}$, see Figure \ref{fig_balls} and Theorem \tref{thm_possible}.

Remark that, given the stochasticity of $\wh s$, the decision rules applied to the subsystems may fail to reflect the nesting in Theorem \tref{thm_possible}. As a result, the properties of both the single and the joint decision rules are of interest. These are discussed in the following section.

\section{Asymptotic properties}\label{sec_asy}
This section discusses the asymptotic properties of the statistics introduced in Section \tref{sec_main}. 

\subsection{Asymptotic properties of squared canonical correlations}\label{sec_asympt}

{Asymptotics is performed as $(T,K)_{seq}\toinf$, which is consistent with $K/T=o(1)$.
Let $d_t$ in \eqref{eq_d} be constructed using any orthonormal \cadlag\ basis of $L^2[0,1]$ and let $1\geq \ml_1\geq\ml_2\geq\dots\geq\ml_p \geq 0$ be the eigenvalues of \textnormal{$\cca(X_t,d_t)$}, see \eqref{eq_BEP}; FGP showed that for $(T,K)_{seq}\toinf$,
\begin{equation}\label{eq_order_eigs}
\ml_i\convp
\left\{
\begin{array}{lll}
1 & $\qquad$ & i=1,\dots,s\\
0 & & i=s+1,\dots,p
\end{array}\right. ,
\end{equation}
which implies
$\myPr(\wh s = s)\conv 1$ as $(T,K)_{seq}\toinf$ for $\wh s$ equal to the max-gap estimator in \eqref{eq_def_shat} or the alternative argmax criterion in \eqref{eq_criteria}.

Moreover, when
$d_t$ in \eqref{eq_d} is constructed using the {\KLB} as in Section \tref{sec_cca_KL}, see also \eqref{eq_KLB}, and $1\geq \ml_1\geq\ml_2\geq\dots\geq\ml_p \geq 0$ are the eigenvalues of \textnormal{$\cca(X_t,d_t)$}, FGP showed that for $(T,K)_{seq}\toinf$,
the $s$ largest eigenvalues $\ml_1\geq\ml_2\geq\dots\geq\ml_s$ satisfy
\begin{equation}\label{eq_K_IG}
K\pi^2\tau^{(s)}\convw\zeta^{(s)},
\end{equation}
where
$\tau^{(s)}:=(1-\ml_s,1-\ml_{s-1},\dots,1-\ml_1)'$, $\zeta^{(s)}:=(\zeta_1,\zeta_2,\dots,\zeta_s)'$, and
$\zeta_1\geq\zeta_2\geq\dots\geq\zeta_s$ are the eigenvalues of
$\left(\int B_1 B_1'\right)^{-1},$
and $B_1(u)$ is a standard Brownian motion of dimension $s$, see the paragraph below \eqref{eq_D_fullrank} in the Appendix.
Additionally,  the $r$ smallest eigenvalues $\ml_{s+1}\geq\ml_{s+2}\geq\dots\geq\ml_p$ satisfy $K\pi^2 (1-\ml_i)\stackrel{p}{\conv}\infty$, $i=s+1,\dots,p$,
which implies,
$K \pi^2 \|\tau^{(i)}\|_h{\conv}\infty
$ for $i>s$.

Observe that \eqref{eq_K_IG} does not depend on nuisance parameters, and that tests based on $K\pi^2\tau^{(s)}$ are asymptotically pivotal.  One consequence of this is that the quantiles $c_{i,h,\eta}$ of $K \pi^2 \|\tau^{(s)}\|_h\convw \|\zeta^{(s)}\|_h$ can be estimated by Monte Carlo simulation.\footnote{Note that these critical values enter Definition \tref{def_s_check} in Section \tref{sec_est_test}.}
This limit distribution
is not invariant to the choice of $L_2[0,1]$ basis.

Some consequences of these results are collected in the following corollary.

\begin{corollary}[Asymptotic properties of the eigenvalues with the {\KLB}]\label{coro_asy_distr_s}
Let $\wh s_h (X_t)$ be the estimator of $s$ based on the test sequence $K \pi^2 \|\tau^{(i)}\|_h$ for \textnormal{$\cca(X_t,d_t)$}, with significance level $\eta$, see Section \tref{sec_est_test};\footnote{Definition \tref{def_s_check} in the Appendix reports the precise definition of the test sequence.}
 then as $(T,K)_{seq}\toinf$ the following holds:
\begin{enumerate}
\item[(i)]  
$\myPr(\wh s _{h}(X_t) = k) \tozero$ for $k>s$.
\item[(ii)] 
$\myPr(\wh s _{h}(X_t) = s) \to 1-\eta$ for $0<s\leq p$ and
$\myPr(\wh s _{h}(X_t) = s) \toone$ for $s=0$;
\item[(iii)]
$\myPr(\wh s _{h}(X_t) = k) \leq \rho_T \to \eta$ for $k<s$.
\end{enumerate}
\end{corollary}

\subsection{Univariate case}
This section contains a novel result, namely  the explicit form of the p.d.f. and c.d.f. of the asymptotic distribution of $K \pi^2 \|\tau^{(s)}\|_h$, $h=1,\infty$, for the special case $s=1$. This limit distribution is related to the well known functional $\int_{0}^{1}B^2(u)\rd u$ of univariate Brownian motion $B(u)$, which has been studied since \cite{CM:45}, see \cite{Ab:93}, \cite{JN:95} and references therein.
The following theorem derives the formulae for the p.d.f. and c.d.f. of \eqref{eq_K_IG}
using these previous results.
\begin{figure}[htbp]
\begin{center}
\includegraphics[width=0.7\textwidth]{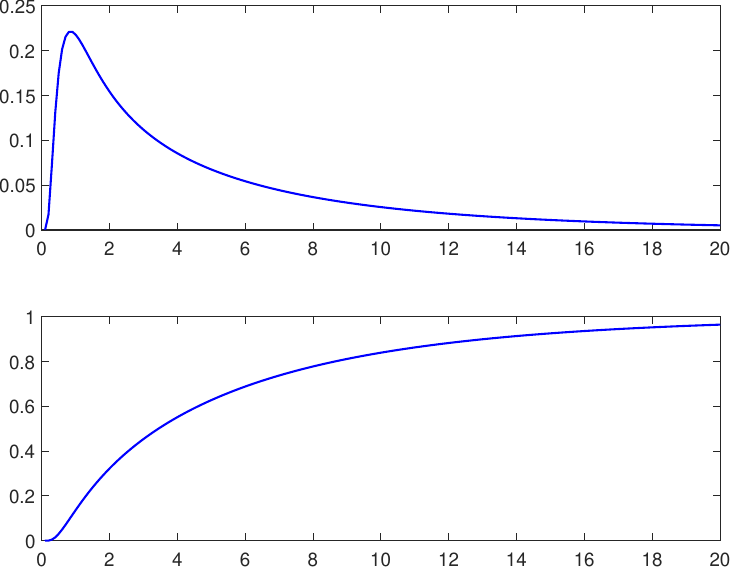}
\end{center}
\caption{p.d.f. and c.d.f. of $\zeta ^{(1)}$ in Theorem \ref{thm_one_dim}.}
\label{fig_one-dim}
\end{figure}

\begin{theorem}[Exact quantiles of $\zeta^{(1)}$]\label{thm_one_dim}
When $s=1$, $\zeta :=\zeta ^{(1)}$ has p.d.f. $f_{\zeta }(z)$ and c.d.f. $F_{\zeta}(z)$ of the form%
\begin{equation*}
f_{\zeta }(z) =\frac1{\sqrt{\pi z}}\sum_{j=0}^{\infty }\eta
_{j}a_{j}\exp\left( -\frac{a_{j}^2}2z\right) ,\quad\qquad 
F_{\zeta }(z) =1-\left(\frac2{\pi }\right) ^{\frac 12%
}\sum_{j=0}^{\infty }\eta_{j}\Gamma \left(\frac 12,\frac{a_{j}^2}2%
z\right) ,\qquad z > 0
\end{equation*}%
where $\eta_{j}:=\binom{-\frac 12}{j}$, $a_{j}:=2j+\frac 12$ and $\Gamma (k,t)=\int_t^{\infty }z^{k-1}e^{-z}\mathrm{d}z$ is the upper incomplete Gamma function.\footnote{$F_{\zeta}(0)=0$.}
Selected quantiles of this distributions are $F^{-1}(0.90)=13.06582$, $F^{-1}(0.95)=17.71180$, $F^{-1}(0.99)=29.01932$.
\end{theorem}

The expression of the p.d.f. can be shown to match e.g. the results in \citet[][Theorem 3]{DNR:03}, who compute
$1-\frac 12\E\left(\int_0^1B(u)^2\rd u\right)^{-1}\simeq - 1.78143$,
which implies $\mathbb{E}(\zeta ^{(1)})\simeq 2(1+1.78143)=5.5629$. This agrees with the analytical expectation that can be derived from the expression $f_{\zeta }(z)$ above, namely
\begin{align*}
\E(\zeta ^{(1)}) &=\int_{0}^{\infty }zf_{\zeta }(z)\rd z
=2^{\frac 12}\sum_{j=0}^{\infty }\eta_{j}a_{j}^{-2}\simeq 5.56291 
.
\end{align*}%

Figure \tref{fig_one-dim} pictures the p.d.f. and c.d.f. in Theorem \ref{thm_one_dim}.\footnote{\textsc{Matlab} and \textsc{R} code to compute the p.d.f., the c.d.f and the quantile functions in Theorem \ref{thm_one_dim} are available at the authors' personal websites.
}

\subsection{Decision rules on $\myH_0$}
This section presents novel asymptotic properties of the joint and single selection criteria, see \eqref{eq_accept_H01}, for the hypothesis $\psi \in \Psi_0$.

\begin{theorem}[Limit behavior of decision rules]\label{theorem_dec_rule}
Let Assumption \tref{ass_DGP} hold
and consider $(T,K)_{seq}\toinf$. The following holds for $\wh s $ equal to the max-gap or the alternative argmax criterion in Section \tref{sec_criteria}:
\begin{enumerate}
\item under the null $\psi \in \Psi_0$, one has
\begin{equation}\label{eq_pconv}
\myPr(\indB = 1)\toone, \qquad
\myPr(\indC= 1)\toone, \qquad
\myPr(\indD = 1)\toone .
\end{equation}
\item under the alternative $\psi \in \Psi_0^{\complement}$, one has
$\myPr( \indB = 0)\toone$ because
$\myPr( \indD = 0)\toone$ irrespectively of the behavior of $\indC$, which is the following: if $\psi \in \Psi_{01}^{\complement}$, one has $\myPr( \indC = 0)\toone$, whereas if
$\psi \in \Psi_{01}\cap \Psi_{02}^{\complement}$, one has $\myPr( \indC = 1)\toone$.
\end{enumerate}
Next let $\wh s$ be equal to the estimator based on the test sequences in Section \tref{sec_est_test}, with significance level $\nu$ $($respectively $\eta)$ associated with the tests involved in $ \indC$  $($respectively $\indD)$; finally define $\varsigma:=0$ for \eqref{eq_H01a}
and $\varsigma:=\eta$ for \eqref{eq_H02a}; 
the following holds:
\begin{enumerate}
\item[(iii)] under the null, $\psi \in \Psi_0$, one has
\begin{align}\label{eq_pconv2}
\myPr(\indB=1) \geq \rho _{1T} \to
1-\nu - \varsigma, \qquad
\myPr( \indC=1)\to
1-\nu,
\qquad 
\myPr( \indD=1)\to 1- \varsigma;
\end{align}
\item[(iv)] under the alternative,
if $\psi \in \Psi_{01}^{\complement}$, one has $\myPr( \indC = 0)\toone$ and $\myPr( \indB= 0)\toone$,
whereas if
$\psi \in \Psi_{01}\cap \Psi_{02}^{\complement}$, one has $\myPr( \indC = 0)\to \nu$ and $\myPr( \indD = 0)\geq \rho_{2T}\to 1- \eta$ which implies
$\myPr( \indB = 0) \geq \rho_{2T} \to 1-\eta$.
\end{enumerate}
\end{theorem}

A few remarks are in order.

\begin{remark}[Properties of criteria]
Both the joint and \mysingle\ decision rules based on the max-gap or the alternative argmax estimator consistently select the hypothesis under the null and reject it under the alternative with limit probability $1$.
This is different from the case of procedures based on test sequences, where the limit probabilities depend on the significance levels $\eta$, $\nu$, as well as to the quantity $\varsigma$, which is
zero in the case of hypothesis \eqref{eq_H01a}.
\end{remark}

\begin{remark}[Non uniformity]
The behavior of the procedures under the alternative
$\psi \in \Psi_{0}^{\complement}$
depends on which part of the alternative holds, in terms of the decomposition in Theorem \tref{thm_possible}. The bounds given in Theorem \tref{theorem_dec_rule}.(iv) are conservative, i.e. they consider the worst case under the alternative. See  empirical power in Table \tref{table_MC}.
\end{remark}

\begin{remark}[Properties of test sequences and selection of $\nu$, $\eta$.]
The criteria based on tests have different asymptotic properties.
Eq. \eqref{eq_pconv2} 
shows how to control the size of the test asymptotically.
Take for instance the case of the single decision rule for \eqref{eq_H01a}
, with $\varsigma=0$; one can choose
$\eta$ as some pre-specified significance level, such as $5\%$, and in this case the single decision rule selected the null correctly with limit probability 1 and reject it when false with limit probability 95\%.

With this setting, the joint criterion makes the correct choice with limit probability at least equal to 95\% under the null and under the alternative.
Similar reasoning applies for the hypothesis \eqref{eq_H02a},
where one can choose
$\nu=\eta=0.025$; in this case the joint (respectively single) criterion makes the correct choice with limit probability 95\% (respectively 97.5\%) under the null and the alternative.
\end{remark}

\section{Simulations}\label{sec_M}

This section reports a Monte Carlo (MC) simulation study; results show that the proposed methods perform well in systems with moderately large cross-sectional and time series dimension.
The DGP is taken from \citet{OW:18} and \citet{BG:22}. The data is generated from
\begin{equation}\label{eq_DGP_OW}
\Delta X_t = \alpha \beta' X_{t-1} + \varepsilon_t, \qquad \varepsilon_t\sim \myiid N(0,I_p),\qquad t=1,\dots,T,\qquad \Delta X_0=X_0=0,
\end{equation}
with $\beta=(0,I_{p-s})'$ and $\alpha=-\beta$, so that $\psi=(I_s,0)'$.
The values of $p,T,K$, and $s$ considered in the MC design are
\begin{equation}\label{eq_DGP_OW_values}
p=20,\qquad T=150,300,\qquad K=\lceil T^{3/4} \rceil, \qquad s=1,10,19,
\end{equation}
thus covering a medium ($p=20$) dimensional system with low to high number of stochastic trends ($s=1,{p}/{2},p-1$). The number of Monte Carlo replications is set to $N=10^4$.

First, the small sample performance of the estimators in detecting the correct number $s$ of stochastic trends in the $p$-dimensional system \eqref{eq_DGP_OW} is considered. The top panel in Table \tref{table_MC} reports the frequency of wrong selection of $s$ based on the different $\wh s$ estimators.
For $T=150$, the max-gap and the sequential test based on $K \pi^2 \|\tau^{(i)}\|_\infty$ behave similarly (upper-left part of the table): for $s=1$ and $s=10$ the frequency of correct detections is close to one, while for $s=19$ it drops down to $70\%$ and $60\%$ respectively.

The alternative argmax estimator performs well only for $s=10$ ($99\%$ of correct detections) and basically always fails when $s=1,19$. The sequential test based on $K \pi^2 \|\tau^{(i)}\|_1$ works well only for $s=1$. This shows that a ratio of $T/p=150/25=7$ does not guarantee a uniformly good performance of all estimators, which is instead achieved for $T=300$ (upper-right part of the table).

\begin{table}[tbp]\small
\begin{tabular}{cc|cccc|cccc}
\hline
\multicolumn{1}{c}{} & \multicolumn{1}{c|}{} & \multicolumn{1}{c}{max-gap} & \multicolumn{1}{c}{alternative} & \multicolumn{1}{c}{$K \pi^2 \|\tau^{(i)}\|_\infty$} & \multicolumn{1}{c|}{$K \pi^2 \|\tau^{(i)}\|_1$} & \multicolumn{1}{c}{max-gap} & \multicolumn{1}{c}{alternative} & \multicolumn{1}{c}{$K \pi^2 \|\tau^{(i)}\|_\infty$} & \multicolumn{1}{c}{$K \pi^2 \|\tau^{(i)}\|_1$} \\
\multicolumn{1}{c}{} & \multicolumn{1}{c|}{} & \multicolumn{1}{c}{} & \multicolumn{1}{c}{argmax} & \multicolumn{1}{c}{} & \multicolumn{1}{c|}{} & \multicolumn{1}{c}{} & \multicolumn{1}{c}{argmax} & \multicolumn{1}{c}{} & \multicolumn{1}{c}{} \\
\hline
\multicolumn{1}{c}{$\myH_0$} & \multicolumn{1}{c|}{$s$} & \multicolumn{4}{c|}{$T=150$} & \multicolumn{4}{c}{$T=300$}\\
\hline
\multicolumn{2}{c|}{}& \multicolumn{8}{c}{Frequency of incorrect selection of $s$}\\
\hline
& 1 &    0 & 0.98 & 0.02 & 0.02 &    0 & 0.06 & 0.04 & 0.04\\
& 10 & 0.04 & 0.01 & 0.01 & 0.66 &    0 &    0 & 0.02 & 0.01\\
& 19 & 0.30 & 0.99 & 0.39 & 0.93 & 0 & 0 & 0.03 & 0.06\\
\hline
\multicolumn{2}{c|}{}& \multicolumn{8}{c}{Empirical size}\\
\hline
& 1 & 0 & 0.95 & 0.02 & 0.02 &    0 & 0.04 & 0.04 & 0.04\\
\eqref{eq_H01} & 10 & 0.03 & 0.01 & 0.01 & 0.58 &    0 &    0 & 0.02 & 0.01\\
& 19 & 0.01 & 0.99 & 0 & 0 &    0 & 0 & 0.03 & 0\\
\hline
& 1 &    0 & 0.98 & 0.04 & 0.04 &    0 & 0.08 & 0.05 & 0.05\\
\eqref{eq_H02} & 10 & 0.02 & 0.02 & 0.05 & 0.46 &    0 &    0 & 0.07 & 0.06\\
& 19 & 0.25 & 0.95 & 0.21 & 0.90 & 0 & 0 & 0.07 & 0.07\\
\hline
\multicolumn{2}{c|}{}& \multicolumn{8}{c}{Empirical power}\\
\hline
& 1 & 1 & 1 & 1 & 1 & 1 & 1 & 1 & 1\\
\eqref{eq_H11}& 10 & 1 & 1 & 1 & 0.98 & 1 & 1 & 1 & 1\\
& 19 & 1 & 1 & 0.99 & 0.96 & 1 & 1 & 1 & 1\\
\hline
& 1 & 1 & 1 & 1 & 1 & 1 & 1 & 1 & 1\\
\eqref{eq_H12}& 10 & 1 & 1 & 1 & 1 & 1 & 1 & 1 & 1\\
& 19 & 1 & 1 & 1 & 1 & 1 & 1 & 1 & 1\\
\hline
\hline
\end{tabular}
\vspace{1em}
\caption{\small Top panel: frequency of incorrect selection of $s$. Middle panel: rejection frequency under the null \eqref{eq_H01} and \eqref{eq_H02}. Bottom panel: rejection frequency under the alternative \eqref{eq_H11} and \eqref{eq_H12}. The sequential tests are conducted at $5\%$ significance level and the number of Monte Carlo replications is set to $N=10^4$.}\label{table_H_s_1}
\label{table_MC}
\end{table}

Next, in order to investigate the small sample performance of the inference procedures in Theorem \tref{theorem_dec_rule} under the null, consider the hypotheses
\begin{align}
\myH_0 \mbox{ of type \eqref{eq_H01a}}&: \:\: \col\psi \subseteq \col (e_1,\dots,e_{p-1}) && (\mbox{i.e.} \col e_p \subseteq \col \beta), \label{eq_H01}\\
\myH_0 \mbox{ of type \eqref{eq_H02a}} & : \:\: \col e_1\subseteq \col\psi  && (\mbox{i.e.} \col \beta \subseteq \col (e_2,\dots,e_p)), \label{eq_H02}
\end{align}
which hold in the DGP. Hypothesis \eqref{eq_H01} is of type \eqref{eq_H01a} with $A=b_\bot=(e_1,\dots,e_{p-1})$ and $A_\bot=b=e_p$, while hypothesis \eqref{eq_H02} is of type \eqref{eq_H02a} with $a=B_\bot=e_1$, $q=1$, and $a_\bot=B=(e_2,\dots,e_p)$. Inference on them is performed via the $z$ statistic in \eqref{eq_accept_H01}, which checks for \eqref{eq_H01} whether (i) $A'X_t$ (the first $p-1=19$ variables) contains $s=19$ stochastic trends and (ii) $A_\bot'X_t$ (the last variable) is stationary. For \eqref{eq_H02} it checks whether (i) $a'X_t$ (the first variable) contains $q=1$ stochastic trends and (ii) $a_\bot'X_t$ (the last $p-1=19$ variables) contains $s-q=19-1=18$ stochastic trends.

The second block of rows in Table \tref{table_MC} (corresponding to \eqref{eq_H01}) reports the frequency of $z = 0$ (i.e. of rejection) under the null, given that \eqref{eq_H01} holds in the DGP, using different estimators $\wh s$. Similarly to above, $T=150$ does not guarantee a uniformly good performance of all estimators, with the max-gap and the sequential test based on $K \pi^2 \|\tau^{(i)}\|_\infty$ performing very well uniformly in $s$, the alternative argmax estimator performing well only for $s=10$ and the sequential test based on $K \pi^2 \|\tau^{(i)}\|_1$ for $s=1$ and $s=19$. For $T=300$ a uniformly good performance of all estimators is instead achieved.

Similar results hold for hypothesis \eqref{eq_H02}, see the third block of rows in Table \tref{table_MC} (corresponding to \eqref{eq_H02}), with a slightly less satisfactory behavior of the alternative argmax estimator and the sequential test based on $K \pi^2 \|\tau^{(i)}\|_1$ when $T=150$.

Finally, in order to investigate the small sample performance of the inference procedures in Theorem \tref{theorem_dec_rule} under the alternative, consider the hypotheses
\begin{align}
\myH_0 \mbox{ of type \eqref{eq_H01a}}&: \:\: \col\psi \subseteq \col (e_2,\dots,e_p) && (\mbox{i.e.} \col e_1 \subseteq \col \beta), \label{eq_H11}\\
\myH_0 \mbox{ of type \eqref{eq_H02a}} & : \:\: \col e_p\subseteq \col\psi  && (\mbox{i.e.} \col \beta \subseteq \col (e_1,\dots,e_{p-1})), \label{eq_H12}
\end{align}
which do not hold in the DGP. Hypothesis \eqref{eq_H11} is of type \eqref{eq_H01a} with $A=b_\bot=(e_2,\dots,e_p)$ and $A_\bot=b=e_1$, while hypothesis \eqref{eq_H12} is of type \eqref{eq_H02a} with $a=B_\bot=e_p$, $q=1$, and $a_\bot=B=(e_1,\dots,e_{p-1})$.

The bottom panel of Table \tref{table_MC} reports the frequency of $z = 0$, i.e. the rejection frequency under the alternative, using different estimators: the empirical power of $z$ in \eqref{eq_accept_H01} appears uniformly good for all of them already for $T=150$.

\section{Empirical application}\label{sec_app}

This section provides an illustration of the methods on a panel of daily exchange rates from January 04, 2022 to August 30, 2024 of the US dollar against 20 World Markets (WM) currencies, downloaded from the Federal Reserve Economic Data (FRED) website, https://fred.stlouisfed.org/. See \citet{OW:19} for a similar dataset and a review of the literature on the topic.

The sample size is $T=667$ and the total number of series is $p=20$, with a ratio $\lfloor T/p \rfloor=33$. Data (in logs and normalized to start at 0) are plotted in Figure \tref{fig_all_XR}, where the 20 WM currencies are grouped into 6 European (EU) exchange rates (Denmark $(DK)$, Eurozone $(Euro)$, Norway $(NK)$, Sweden $(SK)$, Switzerland $(SF)$, United Kingdom $(BP)$) and 14 Non-European (Non-EU) exchange rates (Australia, Brazil, Canada, China, Hong Kong, India, Japan, Malaysia, Mexico, Singapore, South Africa, South Korea, Taiwan, Thailand).

\begin{figure}[htbp]
\includegraphics[width=0.9\textwidth]{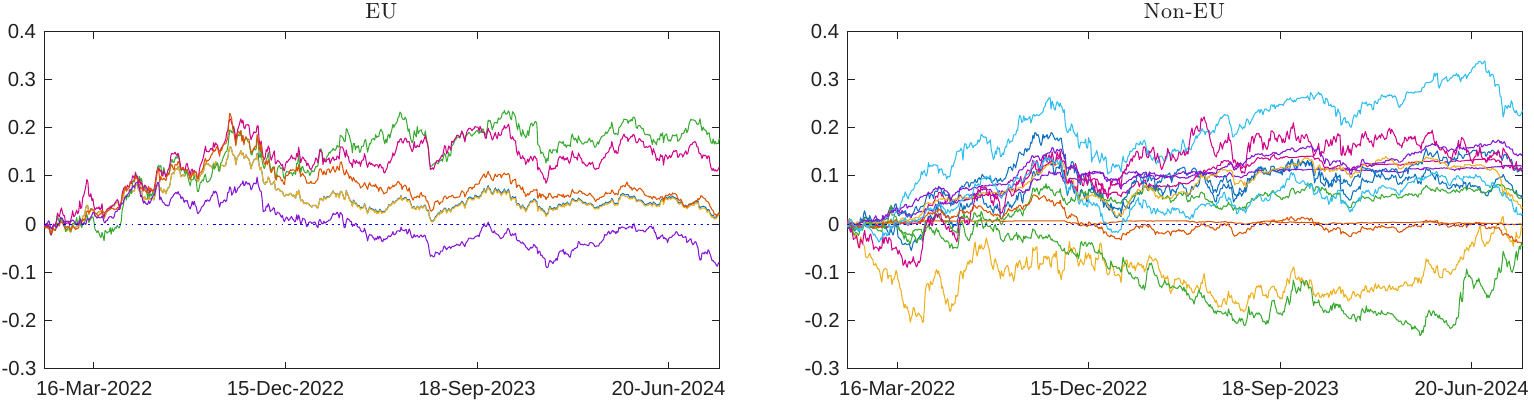}
\caption{\footnotesize Daily exchange rates (in logs and normalized to start at 0) from January 04, 2022 to August 30, 2024 of the 20 WM currencies: 6 EU (left panel) and 14 Non-EU (right panel).}
\label{fig_all_XR}
\end{figure}

\begin{figure}[htbp]
\includegraphics[width=0.9\textwidth]{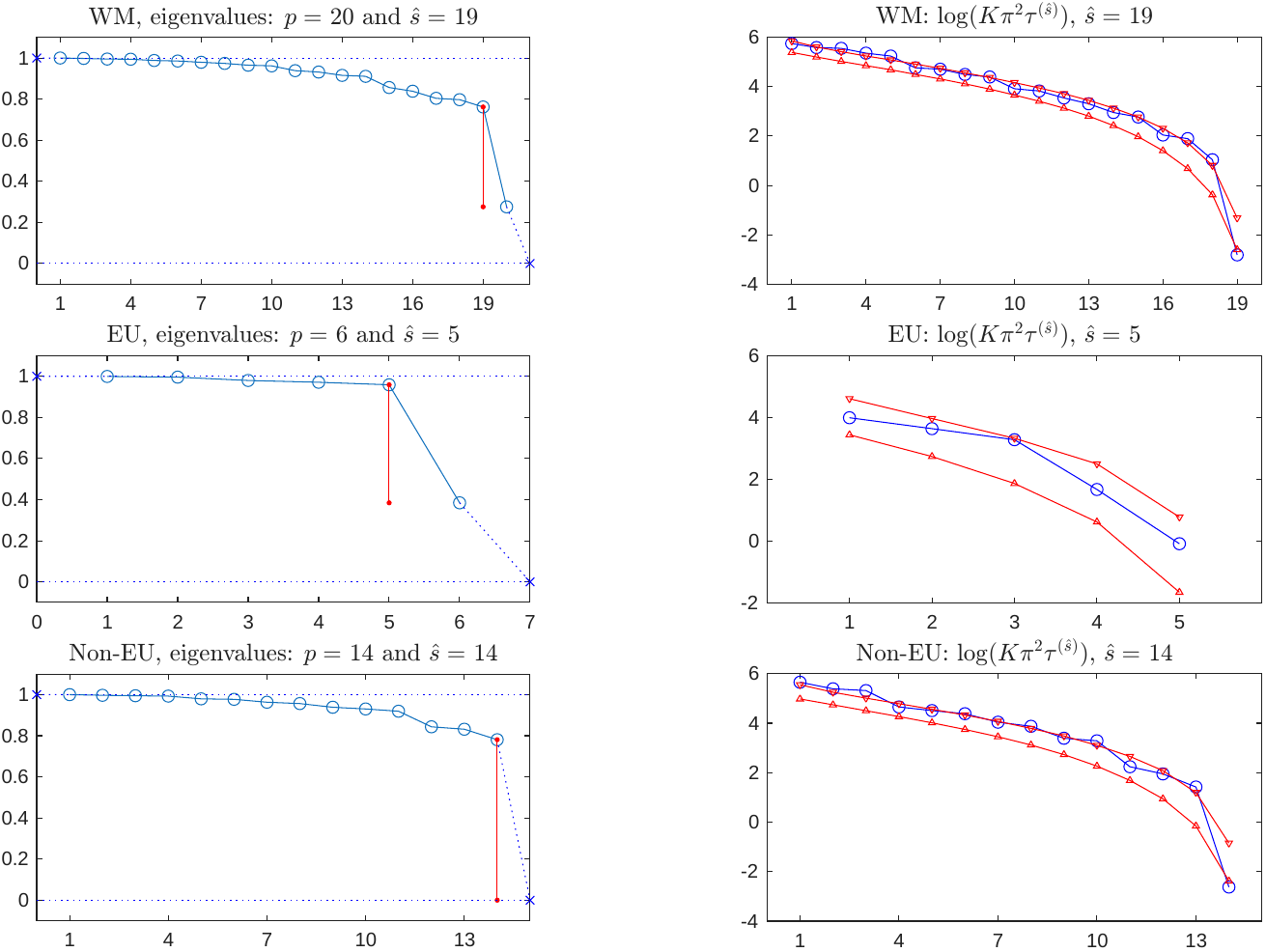}
\caption{\small Analysis of the 20 WM (top panels), the 6 EU (middle panels) and the 14 Non-EU (bottom panels) currencies: eigenvalues, max-gap estimator and $\log(K\pi^2\tau^{(\hat s)})$ for $K=\lceil T^{3/4}\rceil = 132$ and with a $95$\% stripe around $\E(\log \zeta^{(\hat s)})$.}
\label{fig_eig_XR}
\end{figure}

The top panels in Figure \tref{fig_eig_XR} report results for the $p=20$ WM exchange rates $(X_t)$ with $K=\lceil T^{3/4}\rceil = 132$. The first panel shows the eigenvalues of $\cca(X_t,d_t)$ where the first 19 eigenvalues are above 0.75 while $\ml_{20}$ drops down to 0.25 approximately, so that the largest gap $\ml_i-\ml_{i+1}$ is found between $\ml_{19}$ and $\ml_{20}$, for $i=19$. This indicates the presence of $s=19$ stochastic trends, i.e. $r=1$ cointegrating relation, in the $p=20$-dimensional WM system.
The second panel reports the test statistic $K\pi^2\tau^{(\wh s)}$, see \eqref{eq_K_IG}, and the corresponding $95\%$ asymptotic confidence stripe $\mathcal{B}$, see FGP; given that $K\pi^2\tau^{(\wh s)} \in \mathcal{B}$, the model assumptions -- including the selection of $s$ -- appear approximately valid, hence supporting the choice $s=19$.

\begin{table}[tbp]\small
\begin{tabular}{|c||cccc|}
\hline
\hline
\multicolumn{1}{|c||}{Estimate of $s$} & \multicolumn{1}{c}{max-gap} & \multicolumn{1}{c}{alternative argmax} & \multicolumn{1}{c}{$K \pi^2 \|\tau^{(i)}\|_\infty$} & \multicolumn{1}{c|}{$K \pi^2 \|\tau^{(i)}\|_1$} \\
\hline
WM ($p=20$) & 19 & 19 & 19 & 17 \\
EU ($p=6$) & 5 & 5 & 5 & 5 \\
Non-EU ($p=14$) & 14 & 14 & 13 & 12 \\
\hline
\hline
\end{tabular}
\vspace{1em}
\caption{\small Analysis of the 20 WM, the 6 EU and the 14 Non-EU currencies: estimates of $s$ in with the argmax estimators and with the sequential tests at $5\%$ significance level.}
\label{table_MC}
\end{table}

Table \tref{table_MC} reports results for the argmax estimators and the sequential tests at $5\%$ significance level; the different criteria agree, with the exception of the test $K \pi^2 \|\tau^{(i)}\|_1$. Given these overall results, $s$ is fixed at $19$ for the $p=20$-dimensional WM system in the rest of the analysis.

It is next investigated whether the cointegrating relation can be found  in the $6$-dimensional EU system. This is an hypothesis of type \eqref{eq_H02a}, with $a=(0,I_q)'$, $q=14$, and $a_\bot=(I_{p-q},0)'$, $p-q=6$ and it is tested by checking whether (i) $a'X_t$ (the $14$-dimensional Non-EU system) contains $q=14$ stochastic trends and (ii) $a_\bot'X_t$ (the $6$-dimensional EU system) contains $s-q=19-14=5$ stochastic trends, see Theorem \tref{thm_equiv}.

The results are reported in the middle and bottom panels in Figure \tref{fig_eig_XR}: the first 5 eigenvalues of the $6$-dimensional EU system are close to 1 while $\ml_{6}$ drops down to 0.4 approximately, which indicates the presence of $s=5$ stochastic trends, i.e. $r=1$ cointegrating relation, in the $p=6$-dimensional EU system. Hence (ii) is not rejected using the max-gap estimator. 

The $p=14$-dimensional Non-EU system, see the bottom panels in Figure \tref{fig_eig_XR}, displays eigenvalues all above 0.75, and indicates the presence of $s=14$ stochastic trends, i.e. no cointegration. Hence (i) is also not rejected and thus the hypothesis that the cointegrating relation is found in the  $6$-dimensional EU system is not rejected using the max-gap estimator.

The same conclusion, see Table \tref{table_MC}, is found using the alternative argmax estimator. On the contrary, with the sequential tests one rejects the hypothesis because (ii) is not rejected ($\wh s=5$) but (i) is rejected ($\wh s<14$). Given these overall results, in the rest of the analysis $s$ is fixed at $5$ for the $p=6$-dimensional EU system and at $14$ for the $p=14$-dimensional Non-EU system.

In order to investigate the structure of the \cs\ and of the \as, consider the hypothesis that a given currency differential with the Euro in the $6$-dimensional EU system is stationary. This is an hypothesis of type \eqref{eq_H01a} with $b=A_\bot =e_{j}-e_{Euro}$, where $j\in\{DK, NK, SK, SF, BP\}$, and $A=b_\bot=(e_{DK}+e_{Euro},e_{NK},e_{SK},e_{SF},e_{BP})$ for $DK$ and similarly for the other currencies $NK$, $SK$, $SF$, $BP$. This is tested by checking whether (i) $A'X_t$ (the sum of the given currency and the Euro, and the other currencies) contains $s=5$ stochastic trends and (ii) $A_\bot'X_t$ (the given currency differential with the Euro) contains $0$ stochastic trends, see Theorem \tref{thm_equiv}.

The results are reported in Table \tref{table_EU}: for $NK, SK, SF, BP$, (i) is not rejected (because $\wh s=5$) and (ii) is always rejected (because $\wh s=1$); hence the hypothesis that each of those currency differentials is stationary is rejected using any estimator. The same conclusion is found for $DK$ using the argmax estimators. On the contrary, with the sequential tests one does not reject the hypothesis for $DK$ because both (i) and (ii) are not rejected ($\wh s = 5$ in (i) and $\wh s = 0$ in (ii)). Given these overall results, also the $DK$ currency differential with the Euro is considered non-stationary in the rest of the analysis.

\begin{table}[htbp]\small
\begin{tabular}{|c||cccc|}
\hline
\hline
\multicolumn{1}{|c||}{Estimate of $s$ in (i),(ii)} & \multicolumn{1}{c}{max-gap} & \multicolumn{1}{c}{alternative} & \multicolumn{1}{c}{$K \pi^2 \|\tau^{(i)}\|_\infty$} & \multicolumn{1}{c|}{$K \pi^2 \|\tau^{(i)}\|_1$} \\
\multicolumn{1}{|c||}{} & \multicolumn{1}{c}{} & \multicolumn{1}{c}{argmax} & \multicolumn{1}{c}{} & \multicolumn{1}{c|}{} \\
\hline
$DK$ & 5,1 & 5,1 & 5,0 & 5,0 \\
$NK$ & 5,1 & 5,1 & 5,1 & 5,1 \\
$SK$ & 5,1 & 5,1 & 5,1 & 5,1 \\
$SF$ & 5,1 & 5,1 & 5,1 & 5,1 \\
$BP$ & 5,1 & 5,1 & 5,1 & 5,1 \\
\hline
\hline
\end{tabular}
\vspace{1em}
\caption{\small Analysis of the 6 EU currencies: estimates of $s$ in (i) $A'X_t$ and (ii) $A_\bot'X_t$ with the argmax estimators and with the sequential tests at $5\%$ significance level.}
\label{table_EU}
\end{table}

It is next investigated whether the cointegrating relation can be expressed as a linear combination of the $5$ currency differentials with the Euro in the $6$-dimensional EU system. This is an hypothesis of type \eqref{eq_H02a}, with $B=a_\bot=(e_{DK}-e_{Euro},e_{NK}-e_{Euro},e_{SK}-e_{Euro},e_{SF}-e_{Euro},e_{BP}-e_{Euro})$, $a=B_\bot=\iota$, $q=1$, and it is tested by checking whether (i) $a'X_t$ (the sum of the currencies in the EU system) contains $q=1$ stochastic trends and (ii) $a_\bot'X_t$ (the $5$ currency differentials with the Euro) contains $s-q=5-1=4$ stochastic trends, see Theorem \tref{thm_equiv}.

The results are reported in the first row (All differentials) of Table \tref{table_EU_1}: (i) is not rejected (because $\wh s=1$) and (ii) is not rejected (because $\wh s=4$) using any estimator; hence one concludes that the cointegrating relation can be expressed as a linear combination of the $5$ currency differentials with the Euro.

\begin{table}[htbp]\small
\begin{tabular}{|c||cccc|}
\hline
\hline
\multicolumn{1}{|c||}{Estimate of $s$ in (i),(ii)} & \multicolumn{1}{c}{max-gap} & \multicolumn{1}{c}{alternative argmax} & \multicolumn{1}{c}{$K \pi^2 \|\tau^{(i)}\|_\infty$} & \multicolumn{1}{c|}{$K \pi^2 \|\tau^{(i)}\|_1$} \\
\hline
All differentials & 1,4 & 1,4 & 1,4 & 1,4 \\
$DK,NK$ differentials & 4,1 & 4,2 & 4,1 & 4,1 \\
\hline
\hline
\end{tabular}
\vspace{1em}
\caption{\small Analysis of the 6 EU currencies: estimates of $s$ in (i) $a'X_t$ and (ii) $a_\bot'X_t$ with the argmax estimators and with the sequential tests at $5\%$ significance level.}
\label{table_EU_1}
\end{table}

It is finally investigated whether the cointegrating relation can be expressed as a linear combination of only the $DK$ and the $NK$ currency differentials, as found in FGP. This is an hypothesis of type \eqref{eq_H02a}, with $B=a_\bot=(e_{DK}-e_{Euro},e_{NK}-e_{Euro})$, $a=B_\bot=(e_{DK}+e_{Euro}+e_{NK},e_{SK},e_{SF},e_{BP})$, $q=4$, and it is tested by checking whether (i) $a'X_t$ (the sum of $DK$, $NK$ and the Euro, and the other currencies) contains $q=4$ stochastic trends and (ii) $a_\bot'X_t$ (the $DK$ and the $NK$ differentials with the Euro) contains $s-q=5-4=1$ stochastic trends, see Theorem \tref{thm_equiv}.

The results in the second row ($DK,NK$ differentials) of Table \tref{table_EU_1} show that, using the max-gap estimator and the sequential tests, one concludes that the cointegrating relation can be expressed as a linear combination of only the $DK$ and the $NK$ currency differentials; indeed, (i) is not rejected (because $\wh s=4$) and (ii) is not rejected (because $\wh s=1$). On the contrary, with the alternative argmax estimator the hypothesis is rejected because $\wh s=2$ in (ii). Given these overall results, one concludes that the cointegrating relation can be expressed as a linear combination of only the $DK$ and the $NK$ currency differentials.

\section{Conclusions}\label{sec_conc}

This paper proposes a unified semiparametric framework for inference on hypotheses on the \cs\ and \as\ for $I(1)/I(0)$ processes; this includes VARIMA and Dynamic Factor Model processes. 

The equivalence between hypotheses on the \as\ and the number of common stochastic trends in specific subsystems 
is used together with dimensional coherence of the semiparametric approach to define both selection criteria and sequences of tests of hypotheses. In the special one-dimensional case, the limit distribution of interest is found analytically.

It is shown that the selection criteria provide consistent decisions on the hypotheses of interest while test sequences  correctly select the null and alternative with limit probabilities that a depend on the significance level used in the sequence. The properties of the estimators and tests are compared with alternatives, both theoretically and with a Monte Carlo experiment, and results are illustrated on an empirical application to exchange rates. The tools appear useful in discovering properties of the \cs\ and the \as\ in systems with moderately large cross-sectional and time series dimension.

\appendix
\section{Proofs}\label{sec_proofs}

This Appendix contains analytical results and proofs. Some well-known results on the functional central limit theorem and the $L^2$-representation of Brownian motion are first reported.

By the functional central limit theorem, see \citet{PS:92}, the partial sums $T^{-\frac12}\sum_{i=1}^t\varepsilon_i$ in the CT representation \eqref{eq_X} converge weakly to an $n_{\varepsilon}$-dimensional Brownian Motion $W_\varepsilon(u)$, $u\in[0,1]$, with variance $\Omega_\varepsilon>0$; this implies that, with $t =\lfloor Tu\rfloor\in \BN$ and $u\in[0,1]$, one has
\begin{equation}\label{eq_x}
T^{-\frac12}
\left(\begin{array}c
\bar\psi' X_t\\
\beta'\sum_{i=1}^t  X_i
\end{array}
\right)=
T^{-\frac12}\left(\begin{array}c
\kappa'\\
\beta'C_1
\end{array}
\right)\sum_{i=1}^{\lfloor Tu\rfloor}\varepsilon_i +o_p(1)
\underset{T\toinf}{\convw}
\left(\begin{array}c
W_1(u)\\
W_2(u)
\end{array}
\right):= D
W_\varepsilon(u),
\end{equation}
where $\convw$ indicates weak convergence of probability measures on $D_p[0,1]$, $W(u):=( W_1(u)', W_2(u)')'$ is a $p\times 1$ Brownian motion in $u\in[0,1]$ with nonsingular long-run variance $\Omega:=D\Omega_\varepsilon D'$, $D:=(\kappa,C_1'\beta)'$, thanks to Assumption \tref{ass_DGP}.(iii) because
\begin{equation}\label{eq_D_fullrank}
\rank D=
\rank\left(\begin{array}c
\kappa'\\
\beta'C_1
\end{array}
\right) (\bar\kappa,\kappa_\bot)
=\rank\left(\begin{array}{cc}
I_s & 0\\
\beta'C_1\bar\kappa &\beta'C_1\kappa_\bot
\end{array}
\right)=s+\rank\beta'C_1\kappa_\bot.
\end{equation}

Note here that $\beta = \psi _ \bot$ and that Assumption \tref{ass_DGP}.(iii) requires $\psi_\bot'C_1\kappa_\bot$ to be full rank. The $o_p(1)$-term in \eqref{eq_x} is infinitesimal uniformly in $u\in[0,1]$.
It is useful to partition $\Omega$ conformably with $W(u)$ as $\Omega=(\Omega_{ij})_{i,j=1,2}$ and define the standardized version of $W_1(u)$ as $B_1(u):=\Omega_{11}^{-\frac12} W_1(u)$.

Recall, see e.g. \citet{Phi:98}, that any $n$-dimensional standard Brownian motion $B(u)$ admits the representation
\begin{equation}\label{eq_B_L2}
B(u)\simeq\sum_{k=1}^\infty c_k\phi_k(u),\qquad c_k:=\int_0^1B(u)\phi_k(u)\rd u=\nu_k\xi_k,\qquad\nu_k\in \BR_+,\qquad \xi_k\sim N(0,Q),
\end{equation}
where $Q=I_n$, $\{\phi_k(u)\}_{k=1}^\infty$
is an orthonormal basis of $L^2[0,1]$ and $\simeq$ indicates that the series in \eqref{eq_B_L2} is a.s. convergent in the $L^2$ sense to the l.h.s.

In the special case where $(\nu_k^2,\phi_k(u))$, $k=1,2,\dots$, is an eigenvalue-eigenvector pair of the covariance kernel of the standard Brownian motion, i.e. $\nu_k^2\phi_k(u)=\int_0^1\min(u,v)\phi_k(v)\rd v$, \eqref{eq_B_L2} is the {\KL} (KL) representation of $B(u)$, for which one has
\begin{equation}\label{eq_KLB}
\nu_k=\frac1{(k-\frac12)\pi},\qquad\phi_k(u)=\sqrt 2\sin \left(u/\nu_k\right),\qquad\xi_k\:\myiid\:N(0,Q).
\end{equation}
In this case the series in \eqref{eq_B_L2} is a.s. uniformly convergent in $u$ and $\simeq$ is replaced by $=$. In the following a basis $\{\phi_k(u)\}_{k=1}^\infty$ is indicated as the {\KLB} when $\phi_k(u)$ is chosen as in \eqref{eq_KLB}. It follows that the $L^2$-representation of $W(u)$ in \eqref{eq_x} is given by \eqref{eq_B_L2} and \eqref{eq_KLB} with $Q=\Omega$ and $\simeq$ replaced by $=$ when the {\KLB} is employed.

Next, proofs are reported.

\begin{mproof}{\textit{Proof of Proposition} \tref{prop_lin_aggr}}
(a). Aggregation invariance requires $s\geq 1$ (because $X_{t}=I(1)$) and
$\col(v_{1},\dots, v_{p}) \subseteq \col \beta$ with $v_i:=w-e_i$, $i \in \mathcal{I}$. Recall also that $\col(\beta)^\bot = \col \psi$.

Consider the square $p \times p$ matrices $V=(v_{1},\dots, v_{p})=(w-e_1, \dots w - e_p)=w \iota' - I_p$, $B=(w,e_1-e_p,\dots,e_{p-1}-e_{p})$, $C:=(\iota, e_1, \dots, e_{p-1})'$. Observe that $C$ is nonsingular because it is a rearrangement of a triangular matrix with ones on the diagonal. Similarly $CB$ is nonsingular because it is a triangular matrix with ones on the diagonal; this implies that $B$ is full rank as well. Next consider $VB=(0,e_1-e_p,\dots,e_{p-1}-e_{p})$ and hence $\rank V = \rank (VB)  = p-1 $ and $\col(v_{1},\dots, v_{p})=\col (e_1-e_p,\dots,e_{p-1}-e_{p})$. Finally $(\col  (e_1-e_p,\dots,e_{p-1}-e_{p}))^\bot = \col \iota$.
One hence has $\col(v_{1},\dots, v_{p}) \subseteq \col \beta$ if and only if
$\col \psi \subseteq \col \iota$. Because $s\geq 1$, one has $\col \psi = \col \iota$ and one can choose $\psi = \iota$.

(b). Rearrange $X_t$ so as to list regions in $\mathcal{I}_{1}$ first, corresponding to the first $m$ elements of $X_t$. Apply part (a) to $(X_{1t}, \dots X_{mt})'$, and note that the stochastic trends representation for the whole $X_t$ is obtained by padding the loading matrix $\psi$ with zeros for the remaining variables $X_{it}$, $i=m+1,\dots,p$.
\end{mproof}
\smallskip

\begin{mproof}{\textit{Proof of Theorem} \tref{thm_equiv}.}
The statement of the theorem is rewritten as follows for better readability:
\begin{enumerate}
\item[(i)] $\psi =A\theta $ with $\rank \theta = s$ if and
only if \textnormal{(ii.1)} $\rank(\bar{A}'\psi) =s$ and \textnormal{(ii.2)} $\rank(A_\bot'\psi )=0$;
\item[(iii)]$\psi = (a,a_\bot\varphi )$ with $\rank \varphi  = s-q$ if and only if \textnormal{(iv.1)} $\rank(\bar{a}'\psi) =q$ and \textnormal{(iv.2)} $\rank(\bar{a}_\bot'\psi)=s-q$.
\end{enumerate}
Consider each step separately.

\noindent (i) implies (ii.1) and (ii.2): pre-multiply $\psi =A\theta
$ by $(\bar{A},A_\bot)'$ and obtain $s = \rank (\bar{A},A_\bot)'\psi =(\theta',0')'$ which implies (ii.1) and (ii.2).

\noindent (ii.1) and (ii.2) imply (i): by orthogonal projections $\psi =A\bar{A}'\psi +A_\bot\bar{A}_{\bot
}'\psi $; next substitute from (ii.1) and (ii.2) to
find $\psi =A\theta $ with $\theta:=\bar{A}'\psi$ of rank $s$ by (ii.1).

\noindent(iii) implies (iv.1) and (iv.2): pre-multiply $\psi =(a,a_\bot\varphi) $ by $(\bar{a},\bar{a}%
_\bot)'$ and obtain $\bar{a}'\psi =(I_q,0)$ of rank $q$ and $\bar{a}_\bot'\psi
=\varphi$ of rank $s-q$.

\noindent (iv.1) and (iv.2) imply (iii): because $\rank (\bar{a}'\psi)=q$ one can rotate $a$ so that
$\bar{a}'\psi=(I_q,0)$. By orthogonal projections $\psi =a\bar{a}'\psi +a_\bot\bar{a}_{\bot
}'\psi $; next substitute  from (iv.1) and (iv.2) to find $%
\psi =(a,a_\bot\varphi) $ with $\varphi :=\bar{a}_{\bot
}'\psi$ of rank $s-q$.

This completes the proof.
\end{mproof}

\begin{mproof}{\textit{Proof of Theorem} \tref{thm_possible}}
Note that
$\psi \in \Psi_{01}^{\complement}$ means
$\rank\left( H'\psi \right) =:n-i<n$ for some integer $i>0$, and that
$\psi \in \Psi_{02}$ means
$\rank\left( H_\bot'\psi \right) =s-n$. Note that these joint statements would  contradict \eqref{eq_rank_ineq} because this would imply
$s \leq n-i+s-n =s-i$, with $i>0$.
This proves (i) $\Psi_{01}^{\complement} \cap \Psi_{02} = \emptyset$.
In turn this implies  $\Psi_{01}^{\complement} \cap \Psi_{02}^{\complement} = \Psi_{01}^{\complement}$, i.e.
$\Psi_{01}^{\complement} \subset   \Psi_{02}^{\complement}$, and  hence (ii)
$\Psi_{0}^{\complement} = \Psi_{02}^{\complement}$, see Figure \ref{fig_balls}. Finally one finds decomposition (iii) using (i) and (ii), see Figure \ref{fig_balls}.
\end{mproof}

The next definition provides the precise definition of the selection procedure $\wh s$ based on test sequences using $\cca(X_t,d_t)$, see Section \tref{sec_asympt}.

\begin{defn}[Estimators for $s$ based on tests]\label{def_s_check}
The test-based estimator of $s$ at significance level $\eta$ is defined as
\begin{align}\label{eq_def_check}
& \{\wh s  (X_{t}) =j\}:=
\left\{
\begin{array}{lll}
\myA_{p,h,\eta} & & j=p\\
( \cap_{i=j+1}^p \myR_{i,h,\eta} ) \cap \myA_{j,h,\eta}  & & j=1,\dots,p-1\\
\cap_{i=1}^p \myR_{i,h,\eta} & & j=0
\end{array}\right. ,
\\
\nonumber
& \myR_{i,h,\eta}:=\{K \pi^2 \|\tau^{(i)}\|_h >c_{i,h,\eta}\},
\qquad \myA_{i,h,\eta}:=\myR_{i,h,\eta}^\complement=\{K \pi^2 \|\tau^{(i)}\|_h\leq c_{i,h,\eta}\},
\\ \nonumber
&
\tau^{(i)}:=(1-\ml_i,1-\ml_{i-1},\dots,1-\ml_1)',
\end{align}
where $1\geq \ml_1\geq\ml_2\geq\dots\geq\ml_p \geq 0$ are the eigenvalues of $\cca(X_t,d_t)$, $\|\cdot\|_h$ is the $n$-norm for vectors, and $c_{i,h,\eta}$ are appropriate quantiles discussed prior to Corollary \tref{coro_asy_distr_s}.
The corresponding estimator of $r$ is defined as $\wh r_{h}:=p-\wh s_{h}$.
\end{defn}

\begin{mproof}{\textit{Proof of Corollary \tref{coro_asy_distr_s}}} In this proof empty intersections are interpreted as void.

\noindent (i) Assume $s<k$, and observe that $\myPr(( \cap_{i=k+1}^p \myR_{i,h,\eta} ) \cap \myA_{k,h,\eta}) \tozero$ in \eqref{eq_def_check}; this implies
$\myPr(\wh s _{h}(X_t) = k) \tozero$ for $k>s$.

\noindent (ii) For the true $0 < s\leq p$ one has $\myPr(( \cap_{i=s+1}^p \myR_{i,h,\eta} ) \cap \myA_{s,h,\eta}) \to 1 - \eta$ in \eqref{eq_def_check}; this implies $\myPr(\wh s _{h}(X_t) = s) \to 1-\eta$ for $0<s\leq p$. If $s=0$ one has $\myPr(\cap_{i=1}^p \myR_{i,h,\eta})\toone$ and hence
$\myPr(\wh s _{h}(X_t) = s) \toone$ for $s=0$.

\noindent (iii) Assume $s>k$; one has $\myPr(\wh s _{h}(X_t) = k) \leq \myPr(\wh s _{h}(X_t) \in\{s-1 , \dots, 1, 0\} )=: \rho_T \to \eta$ for $k<s$.

\end{mproof}

\begin{mproof}{\textit{Proof of Theorem } \tref{thm_one_dim}}
Let $S=\int_{0}^1B (u)^2\mathrm{d}u$.
By (10) in  \citet{AP:97},
one has
$$
f_{S}(s)=\frac{s^{-\frac{3}2}}{\sqrt{\pi }}%
\sum_{j=0}^{\infty }\eta_{j}a_{j}\exp\left( -\frac{a_{j}^2}{2s}\right),
$$%
where $\eta_{j}:=\binom{-\frac 12}{j}$, $a_{j}:=2j+\frac 12$. Next
define $\zeta :=1/S=g(S)$; by the transformation theorem $%
S=g^{-1}(\zeta )=1/\zeta $, $|\mathrm{d}g^{-1}(\zeta )/\mathrm{d}%
\zeta |=\zeta ^{-2}$, and hence, for $z>0$
$$
f_{\zeta }(z)=f_{S}(z^{-1})z^{-2}=z^{-\frac 12}\frac1{\sqrt{\pi }%
}\sum_{j=0}^{\infty }\eta_{j}a_{j}\exp\left( -\frac{a_{j}^2}2z\right).
$$%
Note also that
\begin{align*}
F_{\zeta }(t) &=1-\int_t^{\infty }f_{\zeta }(z)\mathrm{d}z=1-\frac1{%
\sqrt{\pi }}\sum_{j=0}^{\infty }\eta_{j}a_{j}\int_t^{\infty }z^{-\frac1{%
2}}\exp\left( -\frac{a_{j}^2}2z\right)\mathrm{d}z\\
&=1-\frac1{\sqrt{\pi }}\sum_{j=0}^{\infty }\eta_{j}a_{j}\Gamma \left(
\frac 12,\frac{a_{j}^2}2t\right)\left(\frac{a_{j}^2}2\right) ^{-%
\frac 12}=1-\left(\frac2{\pi }\right) ^{\frac 12}\sum_{j=0}^{\infty
}\eta_{j}\Gamma \left(\frac 12,\frac{a_{j}^2}2t\right),
\end{align*}%
where $\Gamma (k,t)=\int_t^{\infty }z^{k-1} e ^{-z}\mathrm{d}%
z$ is the upper incomplete Gamma function and the following identity has been used, setting $y=az=h(z)$, $%
z=h^{-1}(y)=y/a$, $\mathrm{d}z=a^{-1}\mathrm{d}y$:%
$$
\int_t^{\infty }z^{k-1}e ^{-az}\mathrm{d}%
z=\int_{at}^{\infty }\left(\frac{y}{a}\right) ^{k-1}e ^{-y}
a^{-1}\mathrm{d}y=a^{-k}\int_{at}^{\infty }y^{k-1}e ^{-y}
\mathrm{d}y=a^{-k}\Gamma \left( k,at\right).
$$
\end{mproof}\smallskip
\begin{mproof}{\textit{Proof of Theorem} \tref{theorem_dec_rule}.}
Observe that
\begin{align}\label{eq_inqH0}
& \myPr(\indB=1) \geq 1-\myPr(\indC=0)-\myPr(\indD=0) := \rho_{1T},\\\label{eq_inqH1}
& \myPr(\indB=0) \geq \max(\myPr(\indC=0),\myPr(\indD=0))  := \rho_{2T}.
\end{align}
To see \eqref{eq_inqH0}, define $\setB:= \{\indC = 1\}$, $\setC:=  \{\indD = 1\}$ (and hence $\setB ^{\complement}= \{\indC = 0\}$, $\setC^{\complement}:=  \{\indD = 0\}$) and note that $\myPr(\indB=1)=1-\myPr(\indB=0)$
where
$\{\indB=0\}=(\setB\cap \setC)^{\complement} = \setB^{\complement}\cup \setC^{\complement}$ so that
$\myPr(\indB=0)\leq \myPr(\setB^{\complement})+
\myPr(\setC^{\complement})$ by Boole's inequality, which gives \eqref{eq_inqH0}. For \eqref{eq_inqH1}, observe that
$\myPr(\indB=0) = \myPr(\setB^{\complement} \cup \setC^{\complement})$, from which
\eqref{eq_inqH1} follows by additivity. In the following, eq. \eqref{eq_inqH0} is used when $\psi \in \Psi_0$, \eqref{eq_inqH1}  when $\psi \in \Psi_0^{\complement}$.

Let $\wh s$ be the max-gap or the alternative argmax criterion. If $\psi \in \Psi_0$, then
$\myPr(\setB)\toone$ and
$\myPr(\setC)\toone$ by  \eqref{eq_order_eigs};
this proves
\eqref{eq_pconv} using \eqref{eq_inqH0}.
If $\psi \in  \Psi_{01}^{\complement} $, then
$\myPr(\setB^{\complement})\toone$ and
$\myPr(\setC^{\complement})\toone$ by  \eqref{eq_order_eigs} and Theorem \tref{thm_possible}.(i); this shows
$\myPr(\indB=0)\toone$ by
\eqref{eq_inqH1}.
If $\psi \in  \Psi_{01}\cap\Psi_{02}^{\complement} $
then
$\myPr(\setB)\toone$ and
$\myPr(\setC^{\complement})\toone$; this shows
$\myPr(\indB=0)\toone$ by
\eqref{eq_inqH1}.

Consider now the case of $\wh s$ equal the $\wh s_h$ estimator based on test sequences, $h=1,\infty$.
If $\psi \in \Psi_0$,
then
$\myPr(\setB^{\complement})\to \nu$ and
$\myPr(\setC^{\complement})\to \varsigma$ by Corollary \ref{coro_asy_distr_s}.(ii);
this proves
\eqref{eq_pconv2} using \eqref{eq_inqH0}. Note that $\varsigma=0$ when $H=A$ because $s-n=0$, and the test sequence has limit probability of correct selection equal to 1 when the number of stochastic trends is 0; in case $H=a$, $\varsigma$ equals the
the size of the tests $\eta$ because  $s-q > 0$ (unless $q=s$, in which case hypothesis $H=a$ coincides with $H=A$).

Next assume $\psi \in \Psi_{0}^{\complement} = \Psi_{02}^{\complement}$, with $\rank (H' _ \bot \psi)  = j > s-n$. Observe that
$ \myPr( \indD= 0) \geq \rho_{3T} := \myPr (\wh s _ h (H' _ \bot X_t) = j) \to 1-\eta
$ by Corollary 6.1 (ii).
Next separate two cases, $\psi \in \Psi_{01} \cap \Psi_{02}^{\complement}$ and $\psi \in \Psi_{01}^{\complement}$; when $\psi \in \Psi_{01}\cap\Psi_{02}^{\complement}$
then $\myPr(\indC=0 ) \to \nu$ by Corollary \tref{coro_asy_distr_s}.(ii) and
this, together with $\myPr( \indD= 0) \geq \rho_{3T} \to 1-\eta$,
implies $\myPr( \indB= 0)\geq \rho_{2T} \to 1 - \eta$ by \eqref{eq_inqH1}.
Instead in case $\psi \in \Psi_{01}^{\complement}$ with $\rank (H' \psi) = k < n$, by Corollary \tref{coro_asy_distr_s}.(i) one finds
$\myPr(\indC = 0) \toone $
and hence
$\myPr(\indB = 0) \toone$ by \eqref{eq_inqH1}.
This completes the proof.
\end{mproof}\smallskip

\end{document}